%% file: R7.tex
\documentclass[10pt, twocolumn]{IEEEtran}

\include{micros}

\usepackage[rgb]{xcolor}
\DeclareMathOperator*{\argmin}{arg\,min}
\DeclareMathOperator*{\argmax}{arg\,max}

\usepackage{slashbox}

\usepackage{psfrag}
\usepackage{amssymb}
\usepackage{amsmath}
\usepackage{cite}
\usepackage{graphics}
\usepackage{graphicx}
\usepackage{epsfig}
\usepackage{subfigure}
\usepackage{url}
\usepackage{amscd}
\usepackage{threeparttable}
\usepackage{enumerate}
\begin{document}

\title{Signal-Dependent Performance Analysis of Orthogonal Matching Pursuit \\for Exact Sparse Recovery}

\author{
Jinming~Wen,  Rui Zhang and  Wei Yu, \textit{Fellow, IEEE}

\thanks{This work has been presented in part at IEEE International Conference on Acoustics, Speech, and Signal Processing (ICASSP), May 12-17, 2019, Brighton, UK.}

\thanks{The work of J. Wen was supported by NSFC (Nos. 11871248,61932010,61932011),
the Guangdong Province Universities and Colleges Pearl River Scholar Funded Scheme (2019),
Guangdong Major Project of  Basic and Applied Basic Research (2019B030302008),
the Fundamental Research Funds for the Central Universities (No. 21618329),
a Post-Doctoral Research Fellowship from the Fonds de Recherche
Nature et Technologies and in part by Natural Sciences and Engineering Research Council of Canada;
The work of R. Zhang was supported by Hong Kong Research Grant Council Grant 16306415 and 16317416;
The work of W. Yu was supported by the Natural
Sciences and Engineering Research Council of Canada through the Canada
Research Chairs program.}

\thanks{ J.~Wen  was  with The Edward S. Rogers Sr. Department of Electrical
and Computer Engineering, University of Toronto, Toronto, ON M5S 3G4,
Canada. He is now with the College of Information Science and Technology,
Jinan University, Guangzhou 510632, China (e-mail: jinming.wen@mail.mcgill.ca).}

\thanks{ R.~Zhang was with the department of Mathematics, Hong Kong University of Science and Technology, Hong Kong.
He is now with Huawei Technologies Co., Ltd (e-mail: zhangrui112358@yeah.net).}

\thanks{W. Yu is with The Edward S. Rogers Sr. Department of Electrical
and Computer Engineering, University of Toronto, Toronto, ON M5S 3G4,
Canada (e-mail:weiyu@comm.utoronto.ca).}

}

\maketitle

\begin{abstract}
Exact recovery of $K$-sparse signals $\x\in \mathbb{R}^{n}$ from linear measurements $\y=\A\x$,
where $\A\in \mathbb{R}^{m\times n}$ is a sensing matrix, arises from many applications.
The orthogonal matching pursuit (OMP) algorithm is widely used for reconstructing $\x$ based on $\y$ and $\A$
due to its excellent recovery performance and  high efficiency.
A fundamental question in the performance analysis of OMP is
the characterizations of the probability of exact recovery of $\x$ for random matrix $\A$
and the minimal $m$ to guarantee a target recovery performance.
In many practical applications, in addition to sparsity, $\x$ also has some additional properties
(for example, the nonzero entries of $\x$ independently and identically follow
a Gaussian distribution, or $\x$ has exponentially decaying property).
This paper shows that these properties can be used to refine the answer to the above question.
In this paper, we first show that the prior information of the nonzero entries of $\x$
can be used to provide an upper bound on $\|\x\|_1^2/\|\x\|_2^2$.
Then, we use this upper bound to develop a lower bound on the probability of exact recovery of $\x$ using OMP in $K$ iterations.
Furthermore, we develop a lower bound on the number of measurements $m$ to guarantee that the
exact recovery probability using $K$ iterations of OMP is no smaller than a given target probability.
Finally, we show that when $K=O(\sqrt{\ln n})$, as both $n$ and $K$ go to infinity,
for any $0<\zeta\leq 1/\sqrt{\pi}$, $m=2K\ln (n/\zeta)$ measurements
are sufficient to ensure that the probability of exact recovering any $K$-sparse
$\x$ is no lower than $1-\zeta$ with $K$ iterations of OMP.
This improves the $m=4K\ln (2n/\zeta)$ result of Tropp {\em{et al}}.
For $K$-sparse $\alpha$-strongly decaying signals
and for $K$-sparse $\x$ whose nonzero entries independently and identically follow the
Gaussian distribution, the number of measurements sufficient for exact recovery
with probability no lower than $1-\zeta$ reduces further to
$m=(\sqrt{K}+4\sqrt{\frac{\alpha+1}{\alpha-1}\ln(n/\zeta)})^2$ and asymptotically $m\approx 1.9K\ln (n/\zeta)$, respectively.
\end{abstract}


\section{Introduction}

In many applications, such as sparse activity detection \cite{CheSY18}, we need to reconstruct a $K$-sparse signal $\x$
(i.e., $\x$ has at most $K$ nonzero entries) from linear measurements:
\beq
\label{e:model}
\y=\A\x,
\eeq
where $\A\in \mathbb{R}^{m\times n}$ ($m\ll n$) is a random sensing matrix
with independent and identically distributed (i.i.d.) Gaussian  $\mathcal{N}(0,1/m)$ entries and
$\y\in \mathbb{R}^m$ is a given observation vector.

Numerous sparse recovery algorithms have been developed to recover $\x$
based on $\y$ and $\A$, such as the convex optimization methods
\cite{CanT05,Don06, CanRT06b, Wai09},
nonconvex  optimization methods \cite{FouL09, DauDFG10},
hard thresholding based algorithms \cite{BluD09, Fou11, SheL18} and
greedy algorithms \cite{pati1993orthogonal, DaiM09, NeeT09}.
Among them, greedy algorithms are particularly popular,
especially when $m,n$ and/or $K$ are large.
The orthogonal matching pursuit (OMP) algorithm~\cite{pati1993orthogonal},
which is described as Algorithm~\ref{a:OMP} on the next page,
is a widely used greedy algorithm due to its
low computational complexity and excellent recovery performance \cite{TroG07b}.

A fundamental question in the analysis of OMP is the characterization of its  recoverability.
To this end, numerous works have studied the recovery performance of OMP (see, e.g.,
\cite{Zha11, WanS16, CohDD17, DonH01,Tro04, DavW10, MoS12, LiuT12, WanS12, WenZL13, ChaW14, Mo15, WenZWM17}).
In particular, \cite{TroG07b}  develops a lower bound on the probability of exact recovery
of $K$-sparse signals $\x$ with $K$ iterations of OMP,
and shows that for any fixed $\delta\in (0,0.36)$,
when $K$ and $n$ approach infinity, any $K$-sparse signal $\x$ can be exactly recovered
in $K$ iterations using OMP with probability exceeding $1-2\delta$
if $m\geq (4+\eta)K \ln(n/\delta)$ for any positive number $\eta$.

As OMP is one of the most popular sparse recovery algorithms, to better understand its recover capability,
it is natural to ask whether the lower bound, on the probability of exact
recovery of sparse signals with OMP, developed in \cite{TroG07b}  can be improved.
Further, as measurements may be expensive and/or time consuming in practice,
it is of interest to reduce the necessary number of measurements
for ensuring that the exact recovery probability of OMP is no less than a certain given target probability.

In many practical applications, in addition to sparsity, $\x$ also has some
other properties. For example, in sparse activity detection \cite{CheSY18},
the nonzero entries of $\x$ are assumed to independently and identically follow
the standard Gaussian distribution $\mathcal{N}(0, 1)$.
In speech communication \cite{HabGC09} and  audio source separation \cite{VinBGB14},
$\x$ may have an exponentially decaying property, i.e., $\x$ is a $K$-sparse $\alpha$-strongly-decaying signal which is defined as:
\begin{definition}[\cite{DavW10}] \label{d:alphastr}
Without loss of generality, let the entries of $K$-sparse $\x$ be ordered as
\beq
 \label{e:order}
|x_1|\geq |x_2|\geq\ldots\geq |x_K|\geq 0, \, x_j=0, \mbox{ for } K+1\leq j\leq n.
\eeq
Then $\x$ is called as a $K$-sparse $\alpha$-strongly-decaying signal ($\alpha>1$) if
$
|x_i|\geq \alpha|x_{i+1}|, 1\leq i\leq K-1.
$
\end{definition}

Intuitively, a larger variation in the magnitudes of the nonzero entries of $\x$
would typically lead to a better exact recovery performance of OMP.
In fact, it has been shown that sufficient conditions of exact recovery of
$K$-sparse $\alpha$-strongly-decaying signals with OMP in $K$ iterations are much weaker than those for
general $K$-sparse signals \cite{DavW10, DinCG13, HerDS16, WenZLLT19}.
There are also some works that use the prior distribution of $\x$ to modify the OMP
and analyze its sufficient condition of stable recovery, see, e.g., \cite{WanLRWG18,GeWWX19}.

This paper aims to develop a theoretical framework to capture the dependence of the exact recovery performance of OMP on the disparity in the magnitudes of the nonzero entries of $\x$.
Toward this end, we define the following measure of the disparity
in terms of a function $\phi(t)$:
\beq
\label{e:csk}
\|\x_{\mathcal{S}}\|_{1}^2\leq \phi(|\mathcal{S}|)
\|\x_{\mathcal{S}}\|_{2}^2,\,\forall\,\mathcal{S}\subseteq \Omega,
\eeq
where $\Omega:=\{i|x_i\neq0\}$ denotes
the support of $\x$, $|\mathcal{S}|$ denotes the number of elements of $\mathcal{S}$
and $\phi(t)$ with $0<\phi(t)\leq t$ is a nondecreasing function of $t>0$.
Note that by the Cauchy-Schwarz inequality, \eqref{e:csk} with $\phi(t)=t$
holds for any $K$-sparse signal $\x$.
Furthermore, \eqref{e:csk} with $\phi(t)$  much smaller than $t$ holds for
$\alpha$-strongly-decaying signals (more details are provided in Section \ref{ss:mainpb}).

\begin{algorithm}[t]
\caption{The OMP Algorithm~\cite{pati1993orthogonal}}  \label{a:OMP}
Input: $\y$, $\A$, and stopping rule.\\
Initialize: $k=0, \rr^0=\y, \mathcal{S}_0=\emptyset$.\\
until the stopping rule is met
\begin{algorithmic}[1]
\STATE $k=k+1$,
\STATE $s^k=\argmax\limits_{1\leq i\leq n}|\langle \rr^{k-1},\A_i\rangle|$,
\STATE $\mathcal{S}_k=\mathcal{S}_{k-1}\bigcup\{s^k\}$,
\STATE $\hat{\x}_{\mathcal{S}_k}=\argmin\limits_{\x\in \mathbb{R}^{|\mathcal{S}_k|}}\|\y-\A_{\mathcal{S}_k}\x\|_2$,
\STATE $\rr^k=\y-\A_{\mathcal{S}_k}\hat{\x}_{\mathcal{S}_k}$.
\end{algorithmic}
Output: $\hat{\x}=\argmin\limits_{\x: \text{supp}(\x)=\mathcal{S}_k}\|\y-\A\x\|_2$.
\end{algorithm}

In this paper, we investigate the recovery performance of OMP for recovering $K$-sparse signals
satisfying \eqref{e:csk}. Specifically, our contributions are summarized as follows:
\begin{enumerate}
\item
We develop a lower bound on the probability of exact recovery of any $K$-sparse
signals $\x$ that satisfy \eqref{e:csk}, using $K$-iterations of OMP,
as a function of $\phi(t)$ (see Theorem \ref{t:noiseless}).
Since the bound depends on $\phi(t)$, we develop closed-form expressions of $\phi(t)$ for general $K$-sparse signals,
$K$-sparse $\alpha$-strongly-decaying signals, and $K$-sparse signals with i.i.d.  $\mathcal{N}(0, \sigma^2)$ entries for any $\sigma$
\footnote{This class of signals are called as $K$-sparse Gaussian signals
for short in this paper},
leading to exact lower bounds for these three classes of sparse signals
(see Corollaries \ref{c:regular}-\ref{c:Gauss}).
More exactly, they are respectively, $\phi(t)=t$, $\phi(t)=\frac{(\alpha^t-1)(\alpha+1)}{(\alpha^t+1)(\alpha-1)}$
and $\phi(t)=t$ for $t<\lceil0.95K\rceil$ and $\phi(t)=0.95K$ otherwise.
This part has been presented in a conference paper \cite{WenY19}.
\item
We develop a lower bound on the necessary number of measurements to ensure that the
probability of exact recovery of $K$-sparse signals $\x$, satisfying \eqref{e:csk},
using $K$-iterations of OMP is no smaller than a given target probability
(see Theorem \ref{t:nbofob}).
By using the closed-form expressions of $\phi(t)$ for the  three classes of
sparse signals, the lower bounds on the number of measurements for these
three classes of sparse signals are obtained
(see Corollaries \ref{c:nbofob}-\ref{c:nbofob4}).
We further show that, for any $0<\zeta\leq1/\sqrt{\pi}$,
when $K=O(\sqrt{\ln n})$, as both $n$ and $K$ go to infinity, $m=2K\ln (n/\zeta)$ measurements
are sufficient to ensure that the probability of exact recovering any $K$-sparse
$\x$ is no lower than $1-\zeta$ using $K$ iterations of OMP (see Corollary \ref{c:nbofob}).
This improves the $m=4K\ln (2n/\zeta)$ result of Tropp {\em{et al.}} \cite{TroG07b}.
For $K$-sparse $\alpha$-strongly-decaying signal and for $K$-sparse Gaussian $\x$,
the number of measurements sufficient for exact recovery
with probability no lower than $1-\zeta$ reduces further to
$m=(\sqrt{K}+4\sqrt{\frac{\alpha+1}{\alpha-1}\ln(n/\zeta)})^2$  and asymptotically $m\approx 1.9K\ln (n/\zeta)$,
respectively (see Corollaries \ref{c:nbofob3} and \ref{c:nbofob4}).

\item
Simulations show that the proposed lower bounds are much better than the existing one in \cite{TroG07b},
and the recovery performances of OMP for recovering $K$-sparse
$\alpha$-strongly-decaying and $K$-sparse Gaussian signals are significantly
better than that for recovering $K$-sparse flat signals (i.e., sparse signals with identical magnitude of nonzero entries).

\item Our analysis theoretically explains why the OMP algorithm has better recovery performance
for recovering sparse signals with larger variation in the magnitudes of their nonzero entries.
\end{enumerate}

There are many papers investigate the recovery performance of sparse
recovery algorithms for recovering $\x$ with certain special structure,
such as \cite{DaiM09}, \cite{DavW10}, \cite{DinCG13,HerDS16, WenZLLT19}, \cite{BouFH16}.
By exploiting the special structure, their recovery performances
are improved as shown in these references.
But as far as we know, this paper is the first to use a function $\phi(t)$
to characterize the structure of $\x$,
the probability of exact sparse recovery with OMP
and the necessary number of measurements to ensure a given target recover probability of OMP.
Furthermore, this paper is the first  to give explicit forms of $\phi(t)$
for $K$-sparse $\alpha$-strongly-decaying and $K$-sparse Gaussian signals,
leading to explicit analyses of exact recovery of these two classes of sparse signals with OMP.

Different from the works
(see, e.g., \cite{Zha11, WanS16, CohDD17, Tro04, DavW10, MoS12, LiuT12, WanS12, WenZL13, ChaW14, Mo15, WenZWM17}))
which use the restricted isometry property (RIP) or mutual coherence framework
to study the sufficient condition of exact recovery of any $K$-sparse signal $\x$ for an arbitrary fixed $\A$,
most results of this paper, as in \cite{TroG07b}, study the probability of exact recovering an arbitrary $K$-sparse signal $\x$,
using $K$ iterations of OMP, for randomly chosen $\A$.
Compare to \cite{Zha11, WanS16, CohDD17, Tro04, DavW10, MoS12, LiuT12, WanS12, WenZL13, ChaW14, Mo15, WenZWM17},
our study is more useful from practical applications point of view.
But since this paper assumes that the entries of $\A$
independently and identically follow the $\mathcal{N}(0,1/m)$ distribution,
and Gaussian matrices are not the only ones that satisfy the RIP,
our paper studies the performance of OMP for a smaller class of sensing matrices than
\cite{Zha11, WanS16, CohDD17, Tro04, DavW10, MoS12, LiuT12, WanS12, WenZL13, ChaW14, Mo15, WenZWM17}.
On the other hand, the RIP-based sharp condition given by \cite[Theorem 1]{WenZWM17} combines with the proof of \cite[Theorem 5.2]{BarDDW08}
implies that to ensure any $K$-sparse signal can be exactly recovered by OMP in $K$ iterations,
the necessary number of measurements $m$ needs to satisfy $m=O(K^2\ln n)$,
but if the exact recovery probability is relaxed to $1-\zeta$ for $\zeta\in (0, 0.72)$, then $m =4K \ln(2n/\zeta)$ measurements are sufficient \cite{TroG07b}.
This paper improves this result in showing that (asymptotically) $m= 2K\ln (n/\zeta)$,
$m=(\sqrt{K}+4\sqrt{\frac{\alpha+1}{\alpha-1}\ln(n/\zeta)})^2$ and $m\approx 1.9K\ln (n/\zeta)$
are sufficient to ensure that any $K$-sparse signal,
any $K$-sparse $\alpha$-strongly-decaying signal and any $K$-sparse Gaussian signal, respectively,
can be exactly recovered with OMP in $K$ iterations with probability no lower than $1-\zeta$ for any $\zeta\in (0, 1/\sqrt{\pi}]$.
While our work already improves the lower bound developed in \cite{TroG07b}
on the probability of exact recovery of
any $K$-sparse $\x$, as shown in Sections \ref{s:main} and \ref{s:sim},
the improvement is more significant for $\x$ with smaller $\phi(t)$
(for example, for the case of $K$-sparse Gaussian signals and $\alpha$-strongly-decaying signals).
Hence, our work is more useful in applications, such as sparse activity detection \cite{CheSY18} and
speech communication \cite{HabGC09}, where the exact recovery of $K$-sparse Gaussian
signals  or $\alpha$-strongly-decaying signals is needed.

The rest of the paper is organized as follows.
We present our main results and prove them in Sections~\ref{s:main} and \ref{s:proof}, respectively.
Simulation tests to illustrate our main results are provided in Section \ref{s:sim}.
Finally, we summarize this paper and propose some future research problems in  Section~\ref{s:con}.

\emph{Notation}:
Let $\e_k$ denote the $k$-th column of an identity matrix $\I$.
Denote $\Omega=\text{supp}(\x)$ be the support of $\x$, $|\Omega|$ be the cardinality of $\Omega$
and let $\Omega^c=\{1,2,\ldots ,n\}\setminus \Omega$.
For any set $\mathcal{S}\subseteq \Omega$,
let $\Omega \setminus {\mathcal{S}}=\{i|i\in\Omega,i \notin S\}$ and $\mathcal{S}^c=\{1,2,\ldots ,n\}\setminus {\mathcal{S}}$.
Let $\A_{\mathcal{S}}$ denote the submatrix of $\A$ that  contains only the columns indexed by $\S$.
Similarly, let $\x_{\mathcal{S}}$ denote the subvector of $\x$ that  contains only the entries indexed by $\mathcal{S}$.
For any matrix $\A_{\mathcal{S}}$ of full column-rank, let $\P_{\mathcal{S}}=\A_{\mathcal{S}}(\A_{\mathcal{S}}^\top\A_{\mathcal{S}})^{-1}\A_{\mathcal{S}}^\top$
and $\P^{\bot}_{\mathcal{S}}=\I-\P_{\mathcal{S}}$ denote the projection and orthogonal complement projection onto the column space of $\A_{\mathcal{S}}$,  respectively, where $\A_{\mathcal{S}}^\top$ stands for the transpose of $\A_{\mathcal{S}}$.
Note that we denote $\P_{\mathcal{S}}=\0$ and $\P^{\bot}_{\mathcal{S}}=\I$ when
$\mathcal{S}=\emptyset$.

\section{Main Results}
\label{s:main}

\subsection{Probability of exact recovery}
\label{ss:mainpb}

In the following, we provide a lower bound on the probability that OMP exactly recovers any $K$-sparse
signal $\x$,  which satisfies \eqref{e:csk}, in $K$ iterations.

\begin{theorem}
\label{t:noiseless}
Suppose that in \eqref{e:model}, $\A\in \mathbb{R}^{m\times n}$ is a random matrix
with i.i.d. $\mathcal{N}(0, 1/m)$ entries,
and $\x$ is a $K$-sparse signal that satisfies \eqref{e:csk} for some $\phi(t)$.
Define the event $\mathbb{S}$ as
\beq
\label{e:E}
\mathbb{S}=\{\mbox{OMP exactly recovers $\x$ in $K$ iterations}\}.
\eeq
Denote interval $\mathcal{I}=\left(0,1-\sqrt{\frac{K}{m}}-\sqrt{\frac{2\phi(K)}{m\pi}}\right]$, then
\begin{align}
\label{e:prob}
\mathbb{P}(\mathbb{S})\geq\sup_{\epsilon\in\mathcal{I}}(1-e^{-\frac{\epsilon^2m}{2}})
\prod_{k=1}^{K}\left(1-\frac{e^{-\frac{\eta^2m}{2\phi(k)}}}{\sqrt{\frac{\pi m}{2\phi(k)}}\eta}\right)^{(n-K)},
\end{align}
 where
\beq
\label{e:t}
\eta=1-\sqrt{K/m}-\epsilon.
\eeq
\end{theorem}

\begin{IEEEproof}
See Section \ref{ss:pb}.
\end{IEEEproof}

\begin{remark}
\label{re:prob}
The significance of Theorem \ref{t:noiseless} is summarized as follows:
\begin{enumerate}
\item
Theoretically, Theorem \ref{t:noiseless} characterizes the recovery performance of OMP.
In practical terms, we can use $\eqref{e:prob}$ to give a lower bound on $\mathbb{P}(\mathbb{S})$.
If the lower bound is large, say close to 1, then we are confident to use
the OMP algorithm to do the reconstruction.
From the simulation tests in Section \ref{s:sim}, we can see that the lower bound
is sharp when $m/K$ is relative large.
Hence, if the lower bound is small, say much smaller than 1, then another more effective recovery algorithm
(such as the basis pursuit \cite{CanT05}) may need to be used.

\item
As far as we know, Theorem \ref{t:noiseless} gives the first lower bound on $\mathbb{P}(\mathbb{S})$
by using \eqref{e:csk} and the $K$-sparsity of $\x$.
Note that \cite[Theorem 6]{TroG07b}  also gives a lower bound on $\mathbb{P}(\mathbb{S})$ which is
\begin{align}
\label{e:probtrop}
\mathbb{P}(\mathbb{S})\geq\sup_{\epsilon\in(0,\sqrt{m/K}-1)}&
(1-e^{-\frac{\epsilon^2K}{2}})\nonumber\\
\times&\left(1-e^{-\frac{(\sqrt{m/K}-1-\epsilon)^2}{2}}\right)^{K(n-K)}.
\end{align}
Different from \eqref{e:probtrop} which only uses the $K$-sparsity property of $\x$,
Theorem \ref{t:noiseless} uses not only the sparsity property of $\x$ but also
\eqref{e:csk} to derive the lower bound.
Since the right-hand sides of \eqref{e:prob} and \eqref{e:probtrop} are complicated,
it is difficult to theoretically compare them.
However, simulation tests in Section \ref{ss:sim1} show that
\eqref{e:prob} provides a much sharper lower bound on $\mathbb{P}(\mathbb{S})$ than \eqref{e:probtrop}.
The sharper lower bound is useful for reducing the necessary number of measurements
for a target probability of exact recovery of OMP.
More details on this are provided in Section \ref{ss:noofm}.
Since Theorem \ref{t:noiseless} uses \eqref{e:csk}, there are also major differences between
the proofs of Theorem \ref{t:noiseless} and \cite[Theorem 6]{TroG07b}; for more details, see Section \ref{ss:pb}.

\item
Theorem \ref{t:noiseless} can theoretically explain why the OMP algorithm has better
recoverability for recovering sparse signals with larger variation in the magnitudes of
their nonzero entries.
Specifically, it is not difficult to see that the right-hand side of \eqref{e:prob} becomes larger as $\phi(t)$
(or essentially $\frac{\|\x_{\mathcal{S}}\|_{1}}{\|\x_{\mathcal{S}}\|_{2}}$ (see \eqref{e:csk}))
becomes smaller. By the Cauchy-Schwarz inequality,
$\frac{\|\x_{\mathcal{S}}\|_{1}}{\|\x_{\mathcal{S}}\|_{2}}$
achieves the maximal value of $\sqrt{|\mathcal{S}|}$
when the magnitudes of all the entries of $\x_{\mathcal{S}}$ are the same.
Hence, generally speaking, the probability of the exact recovery of $K$-sparse signals $\x$, whose non-zero entries have identical magnitudes, has the smallest lower bound.
On the other hand, if the variation in the magnitudes of the nonzero entries of $\x$ is large, then $\frac{\|\x_{\mathcal{S}}\|_{1}}{\|\x_{\mathcal{S}}\|_{2}}$
is small, hence the right-hand side of \eqref{e:prob} is large.
Therefore, the probability of exact recovery of this kind of $K$-sparse signals $\x$ is large.
\end{enumerate}
\end{remark}

As \eqref{e:prob} depends on $\phi(t)$, to lower bound $\mathbb{P}(\mathbb{S})$,
we need to know $\phi(t)$. In the following, we give closed-form expressions of $\phi(t)$
for three cases. We begin with the first case where we only know that $\x$ is $K$-sparse.
By the Cauchy-Schwarz inequality, one can see that \eqref{e:csk} holds for $\phi(t)=t$.
Hence, by Theorem \ref{t:noiseless}, we get the following result which gives a lower bound on
$\mathbb{P}(\mathbb{S})$ for any $K$-sparse signal $\x$.
\begin{corollary}
\label{c:regular}
Suppose that $\A\in \mathbb{R}^{m\times n}$ is a random matrix with i.i.d. $\mathcal{N}(0, 1/m)$ entries
and $\x$ is an arbitrary $K$-sparse signal. Then \eqref{e:prob} holds with $\phi(t)=t$.
\end{corollary}

Next, we give a lower bound on $\mathbb{P}(\mathbb{S})$ for $\alpha$-strongly-decaying
signals which is defined in Definition \ref{d:alphastr}.

The following lemma provides a closed-form expression of $\phi(t)$ which ensures that
\eqref{e:csk} holds for $K$-sparse $\alpha$-strongly-decaying signals.
\begin{lemma}
\label{l:decaying}
Let $\x$ be a $K$-sparse $\alpha$-strongly-decaying signal, then \eqref{e:csk} holds with
\begin{align}
\label{e:phit}
\phi(t)=
\begin{cases}
      \frac{(\alpha^t-1)(\alpha+1)}{(\alpha^t+1)(\alpha-1)} & \alpha>1 \\
      \alpha &\alpha=1
   \end{cases}
 ,  \,\; t>0.
\end{align}
\end{lemma}

\begin{IEEEproof}
See Appendix \ref{ss:pfldecay}.
\end{IEEEproof}

\begin{figure}[t]
\centering
\includegraphics[width=3.0in]{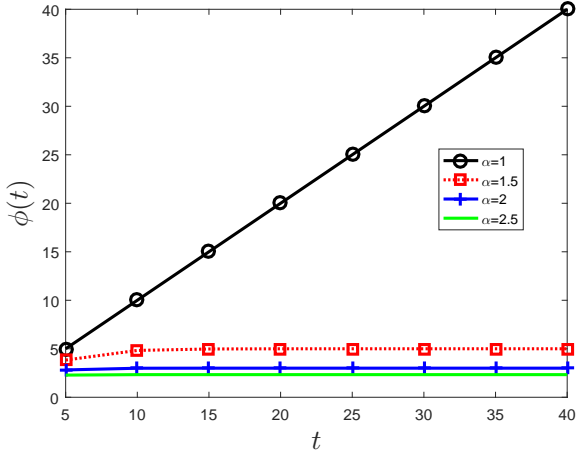}
\caption{$\phi(t)$ defined in \eqref{e:phit} versus $t$ with $\alpha=1, 1.5,2, 2.5$}
\label{f:phi1}
\end{figure}

\begin{figure}[t]
\centering
\includegraphics[width=3.0in]{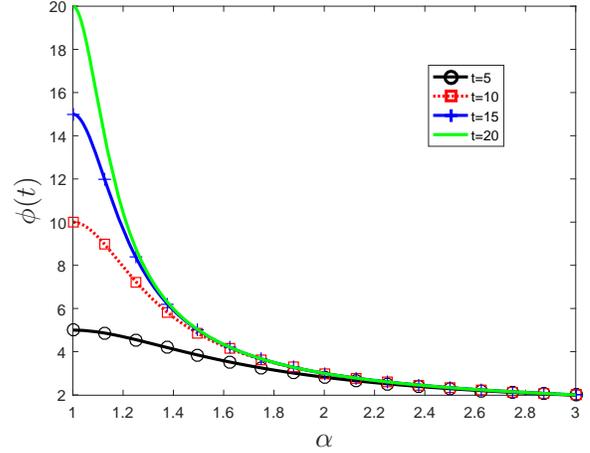}
\caption{$\phi(t)$ defined in \eqref{e:phit} versus $\alpha$ with $t=5, 10, 15, 20$}
\label{f:phi2}
\end{figure}

By the definition of $\alpha$-strongly-decaying signal, $\alpha>1$,
thus $\phi(t)$ for $\alpha>1$ is mainly used in this paper.
Here, $\phi(t)$ for $\alpha=1$ is obtained by taking the limit of
$ \frac{(\alpha^t-1)(\alpha+1)}{(\alpha^t+1)(\alpha-1)}$ with $\alpha$ tends to 1,
and is mainly used for comparing with the general $K$-sparse signal.
Furthermore, by \eqref{e:phit}, $\phi(t)<\frac{\alpha+1}{\alpha-1}$.
Thus, if $\alpha$ is large, say $\alpha\geq 2$, then $\phi(t)\leq 3$ no matter how large $t$ is.
Furthermore, $\phi(t)$ tends to 1 as $t$ tends to infinity,
so $\phi(t)$ can be very close to 1 for any $\alpha>1$ if $t$ is sufficiently large.

To clearly see how large the $\phi(t)$ in \eqref{e:phit} is,
we plot $\phi(t)$ versus $t$ with $\alpha=1, 1.5,2, 2.5$ and $\phi(t)$ versus $\alpha$ with $t=5, 10, 15, 20$ in Figs. \ref{f:phi1} and \ref{f:phi2}.
From  these two figures, we can see that $\phi(t)$ is much smaller than $t$ especially for large $t$ and/or $\alpha$.

By Theorem \ref{t:noiseless} and Lemma \ref{l:decaying}, we get the following corollary.
\begin{corollary}
\label{c:decay}
Suppose that $\A\in \mathbb{R}^{m\times n}$ is a random matrix with i.i.d. $\mathcal{N}(0, 1/m)$ entries
and $\x$ is a $K$-sparse $\alpha$-strongly-decaying signal. Then, \eqref{e:prob} holds with
$\phi(t)$ defined in \eqref{e:phit}.
\end{corollary}

Since $\phi(t)$ defined in \eqref{e:phit} is much smaller than $t$ when $t$ and/or $\alpha$ is large,
the right-hand side of \eqref{e:prob} with $\phi(t)$ defined in \eqref{e:phit} can be much larger than $\phi(t)=t$.
This implies that $\mathbb{P}(\mathbb{S})$ is larger for $\alpha$-strongly-decaying sparse signals
than for flat sparse signals.
Numerical verification of this is given in Section \ref{ss:sim1}.

Finally, we consider the recovery of $K$-sparse Gaussian signals $\x$
with $\x_{\Omega}\sim\mathcal{N}(0, \sigma^2\I)$ for any $\sigma$.
This kind of sparse signals arise from many applications, such as sparse activity
users detection \cite{CheSY18}. If $\x_{\Omega}\sim\mathcal{N}(0, \sigma^2\I)$,
then $\x_{\Omega}/\sigma\sim\mathcal{N}(0, \I)$.
Since $\frac{\|\x_{\mathcal{S}}\|_{1}}{\|\x_{\mathcal{S}}\|_{2}}
=\frac{\|\x_{\mathcal{S}}/\sigma\|_{1}}{\|\x_{\mathcal{S}}/\sigma\|_{2}}$,
to find a function $\phi(t)$ such that \eqref{e:csk} holds for
$K$-sparse $\x$ satisfying $\x_{\Omega}\sim\mathcal{N}(0, \sigma^2\I)$,
we only need to find a function $\phi(t)$ such that \eqref{e:csk} holds for
$K$-sparse $\x$ which satisfies $\x_{\Omega}\sim\mathcal{N}(0, \I)$.

To this end, we introduce the following lemma:
\begin{lemma}
\label{l:ratio}
Suppose that $\u\in \mathbb{R}^{p}$ is a random vector with $\u\sim\mathcal{N}(0, \I)$,
and $0<\mu\leq 1$ is a given constant, then
\begin{align}
\label{e:ratio}
\mathbb{P}(\|\u\|_{1}^{2}\leq \mu p\,\|\u\|_{2}^{2})&\geq 1- \sqrt{1+\gamma}e^{\frac{1}{6}} \nonumber\\
&\times \left(\frac{2.775\gamma^{\gamma}e^{-\frac{\gamma}{2}}(1+\gamma)^{-\frac{1+\gamma}{2}}}{\mu^{\gamma/2}}\right)^{p}
\end{align}
for any $\gamma>0$.
In particular, when $\mu=0.95$, we have
\begin{align}
\label{e:ratio1}
\mathbb{P}(\|\u\|_{1}^{2}\leq 0.95p\,\|\u\|_{2}^{2})&\geq 1-1.87\times0.796^{p}.
\end{align}
\end{lemma}

\begin{IEEEproof}
See Appendix \ref{ss:pflratio}.
\end{IEEEproof}

Fig. \ref{f:rand2} shows the empirical probability and the lower bound given by \eqref{e:ratio1} on
$\|\u\|_{1}^{2}\leq 0.95p\,\|\u\|_{2}^{2}$ for $p=3,4,\ldots,50$ over 50000 realizations
of $\u\in \mathbb{R}^p\sim\mathcal{N}(0, \I)$.
From Fig. \ref{f:rand2}, one can see that although \eqref{e:ratio1} is not
a sharp lower bound for small $p$, it is sharp for large $p$.
\begin{figure}[t]
\centering
\includegraphics[width=3.2in]{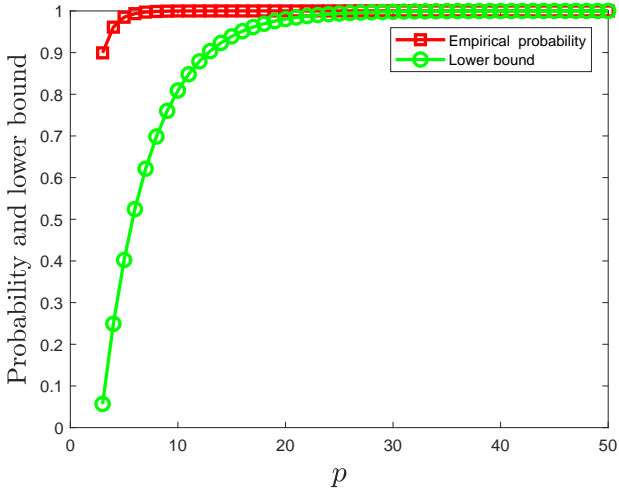}
\caption{The empirical probability and lower bound on $\|\u\|_{1}^2/\|\u\|_{2}^2\leq 0.95 p$
for $p=3,4,\ldots,50$ over 50000 realizations of
$\u\in \mathbb{R}^{p}\sim\mathcal{N}(0, \I)$}
\label{f:rand2}
\end{figure}

By the Cauchy-Schwartz inequality,
$\|\u\|_{1}^{2}\leq t\,\|\u\|_{2}^{2}$ holds for any $\u\in \mathbb{R}^{t}$.
Hence, for simplicity, we define
\begin{align}
\label{e:phit2}
\phi(t)=
\begin{cases}
      0.95K & \lceil0.95K\rceil\leq t\leq K\\
      t & t<\lceil0.95K\rceil
   \end{cases},
\end{align}
where for any $x\in \mathbb{R}$, $\lceil x\rceil$ denotes the smallest integer
that is not smaller than $x$.

The following lemma shows that \eqref{e:csk} holds with $\phi(t)$ being defined in
\eqref{e:phit2} with high probability.

\begin{lemma}
\label{l:Gaussian}
Suppose that $K$-sparse $\x$ satisfies $\x_{\Omega}\sim\mathcal{N}(0, \sigma^2\I)$
for any $\sigma$. Denote $\nu$ as the probability that \eqref{e:csk} holds
with $\phi(t)$ being defined in \eqref{e:phit2}, then
\beq
\label{e:pGaussian}
\nu\geq 1-\frac{3.614}{\sqrt{K}}\times0.981^{\lceil0.95K\rceil}.
\eeq
\end{lemma}

\begin{IEEEproof}
See Appendix \ref{ss:Gaussian}.
\end{IEEEproof}

Fig. \ref{f:phi_LBD} shows the lower bound on $\nu$ given by \eqref{e:pGaussian}.
From Fig. \ref{f:phi_LBD}, one can see that the lower bound is sharp for large
$K$ since it is approximately 1, but it is not sharp for small $K$.
We think this is because the lower bound presented in \eqref{e:ratio1}
is not sharp for small $p$.

\begin{figure}[t]
\centering
\includegraphics[width=3.2in]{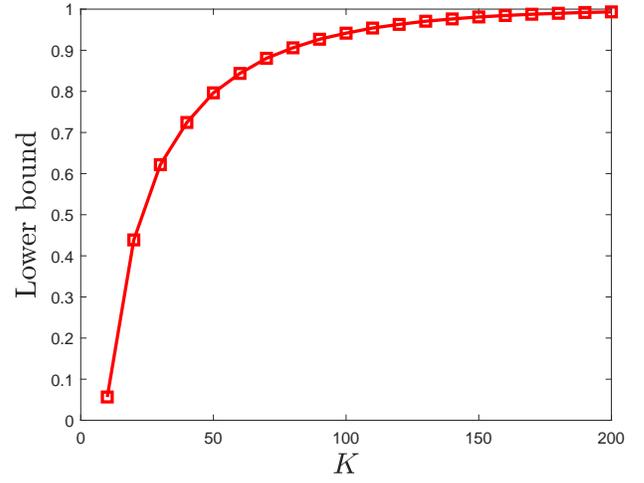}
\caption{The right-hand side of \eqref{e:pGaussian} for $K=10:10:200$}
\label{f:phi_LBD}
\end{figure}

By Lemma \ref{l:Gaussian}, we get the following corollary from Theorem \ref{t:noiseless}:
\begin{corollary}
\label{c:Gauss}
Suppose that $\A\in \mathbb{R}^{m\times n}$ is a random matrix with i.i.d. $\mathcal{N}(0, 1/m)$ entries
and $\x$ is a $K$-sparse signal satisfying $\x_{\Omega}\sim\mathcal{N}(0, \sigma^2\I)$ for
any given $\sigma$.
Then \eqref{e:prob} holds with $\phi(t)$ defined in \eqref{e:phit2} with probability
$\nu$ satisfying \eqref{e:pGaussian}.
\end{corollary}

\begin{remark}
\label{re:gauss}
Although $\phi(t)$ being defined in \eqref{e:phit2} is close to $\phi(t)=t$,
we think a (much) smaller $\phi(t)$ can be found if we can find a sharper
lower bound on $\mathbb{P}(\|\u\|_{1}^{2}\leq \mu p\,\|\u\|_{2}^{2})$ than \eqref{e:ratio}.
Hence, although the lower bound on the probability of exact recovery of
Gaussian sparse signals is close to that for flat sparse signals.
The true probability of the former is much larger than that for the latter.
More details on this is given in Section \ref{ss:sim1}.
\end{remark}

\subsection{Necessary number of measurements}
\label{ss:noofm}

By using Theorem \ref{t:noiseless}, we can prove the following theorem which gives a lower bound on
the necessary number of measurements $m$ to ensure that the OMP algorithm can exactly recover any $K$-sparse vector $\x$ with high probability.

\begin{theorem}
\label{t:nbofob}
For any fixed $K, n\geq2$ and $\delta\in (0,1)$,
let $\beta$ and $m$ respectively satisfy
\begin{align}
\label{e:alpha}
\beta=\max\left\{1, \log_n \frac{(n-K)\sum_{k=1}^{K}\phi(k)}{\phi(K)\sqrt{\ln(n/\delta)}}\right\}
\end{align}
and
\begin{align}
\label{e:m}
m\geq\left(\sqrt{\frac{2\beta\,\phi(K)}{K}}+\sqrt{\frac{1}{\ln(n/\delta)}}+\sqrt{\frac{2}{K}}\right)^2K\ln(n/\delta).
\end{align}
Suppose that  $\A\in \mathbb{R}^{m\times n}$ is a random matrix with i.i.d. $\mathcal{N}(0, 1/m)$ entries
and $\x$ is a $K$-sparse signal satisfying \eqref{e:csk} for some $\phi(t)$.
Then
\begin{align}
\label{e:prob2}
\mathbb{P}(\mathbb{S})\geq1-\left(\frac{\delta}{n}+\frac{\delta^{\beta}}{\sqrt{\pi\beta}}\right),
\end{align}
where event $\mathbb{S}$ is defined in \eqref{e:E}.
\end{theorem}

\begin{IEEEproof}
See Section \ref{ss:pfnm}.
\end{IEEEproof}

By Theorem \ref{t:nbofob}, we get the following corollary.
\begin{corollary}
\label{c:nbofobcoro}
For any fixed $K, n\geq2$ and $\zeta\in (0,1/\sqrt{\pi}]$,
let
\beq
\label{e:delta}
\delta=\frac{n\sqrt{\pi}\zeta}{n+\sqrt{\pi}},
\eeq
$\beta$ and $m$ respectively satisfy \eqref{e:alpha}
and \eqref{e:m}.
Suppose that  $\A\in \mathbb{R}^{m\times n}$ is a random matrix with  i.i.d. $\mathcal{N}(0, 1/m)$
entries and $\x$ is a $K$-sparse signal satisfying \eqref{e:csk} for some $\phi(t)$. Then
\begin{align}
\label{e:prob3}
\mathbb{P}(\mathbb{S})\geq1-\zeta,
\end{align}
where event $\mathbb{S}$ is defined in \eqref{e:E}.
\end{corollary}

\begin{IEEEproof}
Since $\zeta\leq \frac{1}{\sqrt{\pi}}$, by \eqref{e:delta},
we have $\delta<1$.
By \eqref{e:alpha}, $\beta\geq 1$, therefore
\[
\frac{\delta}{n}+\frac{\delta^{\beta}}{\sqrt{\pi\beta}} \leq\frac{\delta}{n}+\frac{\delta}{\sqrt{\pi}}=\zeta.
\]
Therefore, by \eqref{e:prob2}, \eqref{e:prob3} holds.
\end{IEEEproof}

\begin{remark}
\label{re:necessarym}

The significance of Theorem \ref{t:nbofob} and Corollary \ref{c:nbofobcoro} are summarized as follows:

\begin{enumerate}
\item
Corollary \ref{c:nbofobcoro} provides a lower bound on the necessary number of measurements $m$
to guarantee that the probability of the OMP algorithm exactly recovers any $K$-sparse signals $\x$,
that satisfies \eqref{e:csk} for some $\phi(t)$, is no smaller than a given probability.
This is important in many practical applications, where the cost of obtaining the measurements can be large.

\item As far as we know, Theorem \ref{t:nbofob} gives the first lower bound on $m$
which ensures exact recovery with high probability by using \eqref{e:csk}.
Similar bound  has been derived in \cite[Corollary 7]{TroG07b}.
However, since Theorem \ref{t:nbofob} uses not only the sparsity property of $\x$ but also \eqref{e:csk},
while \cite[Corollary 7]{TroG07b}  uses the sparsity property of $\x$ only,
Theorem \ref{t:nbofob} gives a sharper lower bound on $m$ than \cite[Corollary 7]{TroG07b}.
More details on the comparison of the two lower bounds
are contained in Appendix \ref{ss:comp} and the simulation results in Section \ref{ss:sim2}.
Since Theorem \ref{t:nbofob} uses \eqref{e:csk}, the main idea of the proof of Theorem \ref{t:nbofob}
is different from that of \cite[Corollary 7]{TroG07b};
more details are refer to Section \ref{ss:pfnm}.

\item
Corollary \ref{c:nbofobcoro} shows that at a fixed recovery probability,
the necessary number of measurements $m$ becomes smaller as the variation
of the magnitudes of their nonzero entries becomes larger.
Indeed, as explained in the third item of Remark \ref{re:prob},
$\phi(t)$ becomes smaller when the variation in the magnitudes of the nonzero entries of sparse signals becomes larger.
As a result, the necessary number of measurements $m$ tends to be smaller according to \eqref{e:m}.
\end{enumerate}
\end{remark}

By Corollary \ref{c:nbofobcoro}, we can obtain the following corollary.

\begin{corollary}
\label{c:nbofob}
For any fixed $K, n\geq2$ and $\zeta\in (0,1/\sqrt{\pi}]$,
let $\delta=\frac{n\sqrt{\pi}\zeta}{n+\sqrt{\pi}}$,
$\A\in \mathbb{R}^{m\times n}$ be a random matrix with  i.i.d. $\mathcal{N}(0, 1/m)$
entries and $\x$ be a $K$-sparse signal.
If  $\beta$ and $m$ respectively satisfy
\begin{align}
\label{e:alpha2}
\beta&=\max\left\{1, \log_n\frac{(n-K)(K+1)}{2\sqrt{\ln(n/\delta)}}\right\},\\
m&\geq\left(\sqrt{2\beta}+\sqrt{\frac{1}{\ln(n/\delta)}}+\sqrt{\frac{2}{K}}\right)^2K\ln(n/\delta).
\label{e:m2}
\end{align}
Then
\begin{align}
\label{e:prob31}
\mathbb{P}(\mathbb{S})\geq1-\zeta,
\end{align}
where event $\mathbb{S}$ is defined in \eqref{e:E}.
In particular, \eqref{e:prob31} holds in the asymptotic regime in which
 both $K$ and $n$ tend to infinity as $K=O(\sqrt{\ln n})$ and $m\geq2K\ln(n/\zeta)$.
\end{corollary}

\begin{IEEEproof}
By the Cauchy-Schwartz inequality, one can see that \eqref{e:csk} holds for $\phi(t)=t$,
thus by Corollary \ref{c:nbofobcoro}, \eqref{e:prob31} holds if \eqref{e:alpha2} and \eqref{e:m2} hold.
Hence, in the following, we prove the second part of Corollary \ref{c:nbofob}.

By some calculations,  we have
\beqnn
\lim_{K,n\rightarrow\infty}\left(\sqrt{2\beta}+\sqrt{\frac{1}{\ln(n/\delta)}}+\sqrt{\frac{2}{K}}\right)^2
=2\beta.
\eeqnn
Therefore, when both $K$ and $n$ tend to infinity, the right-hand side of \eqref{e:m2}
tends to $2\beta K\ln(n/\delta)$.
Furthermore, since
\begin{align}
\label{e:mappro}
\log_n\frac{(n-K)(K+1)}{2\sqrt{\ln(n/\delta)}}
\leq &\log_n(n-k)+\log_n\frac{K}{\sqrt{\ln(n/\delta)}}\nonumber\\
<&1+\log_n\frac{K}{\sqrt{\ln(n/\delta)}},
\end{align}
$\lim\limits_{K,n\rightarrow\infty}\beta=1$  if $K=O(\sqrt{\ln n})$.
Then, by \eqref{e:delta}, when  both $K$ and $n$ tend to infinity as $K=O(\sqrt{\ln n})$,
$m\geq 2K\ln(n/\zeta)$ is sufficient to ensure \eqref{e:prob31} holds.
\end{IEEEproof}

Note that, by \eqref{e:mappro}, one can see that $\beta$ is very close to 1 if  $\log_n\frac{K}{\sqrt{\ln(n/\delta)}}$
is very close to 0. For example, when $K=10, n=1000$ and $\delta=0.01$,
$\log_n\frac{K}{\sqrt{\ln(n/\delta)}}<0.1565$ and $\beta<1.069$.

Corollary \ref{c:nbofob} shows that $m=2K\ln(n/\zeta)$ measurements are sufficient to  guarantee that
the probability of exact recovering any $K$-sparse signal using $K$ iterations of OMP is no lower than $1-\zeta$
when both $K$ and $n$ tend to infinity as  $K=O(\sqrt{\ln n})$.
This improves Tropp {\em{et al.}}'s result which requires $m\approx4K\ln(2n/\zeta)$ measurements \cite[Corollary 7]{TroG07b};
for more details, see the comparison results in Appendix \ref{ss:comp}.

By Corollary \ref{c:nbofobcoro} and Lemma \ref{l:decaying}, we also get the following corollary.

\begin{corollary}
\label{c:nbofob3}
For any fixed $K, n\geq2$ and $\zeta\in (0,1/\sqrt{\pi}]$,
let $\delta=\frac{n\sqrt{\pi}\zeta}{n+\sqrt{\pi}}$,
$\A\in \mathbb{R}^{m\times n}$ be a random matrix with  i.i.d. $\mathcal{N}(0, 1/m)$
entries and $\x$ be a $K$-sparse $\alpha$-strongly-decaying signal.
If $m\geq (\sqrt{K}+4\sqrt{\frac{\alpha+1}{\alpha-1}\ln(n/\zeta)})^2$, then
\begin{align}
\label{e:prob32}
\mathbb{P}(\mathbb{S})\geq1-\zeta,
\end{align}
where event $\mathbb{S}$ is defined in \eqref{e:E}.
\end{corollary}

\begin{IEEEproof}
By Corollary \ref{c:nbofobcoro} and Lemma \ref{l:decaying},
one can see that if $\beta$ and $m$ respectively satisfy \eqref{e:alpha} and \eqref{e:m} with $\phi(t)$ defined in \eqref{e:phit},
then \eqref{e:prob32} holds.
By \eqref{e:alpha}, one can see that $\beta\leq 2$.
Furthermore, by \eqref{e:phit}, $\phi(K)<\frac{\alpha+1}{\alpha-1}$ for any $K> 1$.
Hence, the right-hand side of \eqref{e:m} is not larger than $(\sqrt{K}+4\sqrt{\frac{\alpha+1}{\alpha-1}\ln(n/\delta)})^2$.
Thus, by \eqref{e:delta}, $m\geq(\sqrt{K}+4\sqrt{\frac{\alpha+1}{\alpha-1}\ln(n/\zeta)})^2$  ensures \eqref{e:prob32} holds.
\end{IEEEproof}

Corollary \ref{c:nbofob3} shows that $m=(\sqrt{K}+4\sqrt{\frac{\alpha+1}{\alpha-1}\ln(n/\zeta)})^2$ measurements are sufficient to ensure
that the probability of exact recovering any $K$ $\alpha$-strongly-decaying signal is no lower than $1-\zeta$,
this significantly improves the requirement $m\approx4K\ln(2n/\zeta)$ in \cite{TroG07b}.
Therefore, the necessary number of measurements $m$ for recovering $\alpha$-strongly-decaying sparse signals can be much smaller than that for recovering flat sparse signals.
More details are referred to Section \ref{ss:sim2}.

By Corollary \ref{c:nbofobcoro} and Lemma \ref{l:Gaussian}, we get the following corollary.

\begin{corollary}
\label{c:nbofob4}
For any fixed $K, n\geq2$ and $\zeta\in (0,1/\sqrt{\pi}]$,
let $\delta=\frac{n\sqrt{\pi}\zeta}{n+\sqrt{\pi}}$,
$\A\in \mathbb{R}^{m\times n}$ be a random matrix with  i.i.d. $\mathcal{N}(0, 1/m)$
entries and $\x$ be a $K$-sparse signal satisfying $\x_{\Omega}\sim\mathcal{N}(0, \sigma^2\I)$ for certain $\sigma$.
If $K=O(\sqrt{\ln n})$, when  both $K$ and $n$ tend to infinity, and $m\geq 1.9K\ln(n/\zeta)$,
then
\beq
\label{e:PSLBD}
\mathbb{P}(\mathbb{S})\geq \nu(1-\zeta),
\eeq
where event $\mathbb{S}$ is defined in \eqref{e:E} and $\nu$ satisfies \eqref{e:pGaussian}.
Note that the probability here is over the randomness of both $\x$ and $\A$.
\end{corollary}

\begin{IEEEproof}
 By \eqref{e:phit2}, we can show that
\[
\lim_{K,n\rightarrow\infty}
\left(\sqrt{\frac{2\beta\phi(K)}{K}}+\sqrt{\frac{1}{\ln(n/\delta)}}+\sqrt{\frac{2}{K}}\right)^2
\leq 1.9\beta.
\]
Hence,  when both $K$ and $n$ tend to infinity, the right-hand side of \eqref{e:m} tends to $1.9\beta K\ln(n/\delta)$.
Furthermore, by \eqref{e:phit2}, we have
\begin{align*}
\log_n \frac{(n-K)\sum_{k=1}^{K}\phi(k)}{\phi(K)\sqrt{\ln(n/\delta)}}
\leq &\log_n \frac{(n-K)(K+1)}{1.9\sqrt{\ln(n/\delta)}}\\
< & 1+\log_n \frac{K+1}{1.9\sqrt{\ln(n/\delta)}}.
\end{align*}
Hence, if both $K$ and $n$ go to infinity as $K=O(\sqrt{\ln n})$, then $\lim\limits_{K, n\rightarrow\infty}\beta=1$.
Hence, according to \eqref{e:delta}, $m\geq1.9K\ln(n/\zeta)$ guarantees \eqref{e:PSLBD} holds.
\end{IEEEproof}

Note that in the asymptotic regime as $K$ goes to infinity (see \eqref{e:pGaussian} and Fig. \ref{f:phi_LBD}), $\nu$ approaches 1.
Thus, Corollary  \ref {c:nbofob4} significantly outperforms the requirement $m\approx4K\ln(2n/\zeta)$ developed in \cite{TroG07b}.
As stated in Remark \ref{re:gauss}, we believe that a $\phi(t)$ which is (much) smaller
than the $\phi(t)$ defined in \eqref{e:phit2} exists, so we believe that
the necessary number of measurements $m$ can be even smaller than $1.9K\ln(n/\zeta)$.

\section{Proofs}
\label{s:proof}

In this section, we prove Theorems \ref{t:noiseless} and \ref{t:nbofob} in Section \ref{s:main}.

\subsection{Useful lemmas}
\label{ss:ul}

To prove Theorems \ref{t:noiseless} and \ref{t:nbofob}, we need to introduce  four useful lemmas.
We begin with the first lemma which characterizes the condition that ensures the $(k+1)$-th
iteration of OMP can find an index in the support $\Omega$ of the $K$-sparse signal $\x$
under the condition that the first $k$ iterations of OMP finds an index in $\Omega$ in each iteration.

\begin{lemma}
\label{l:main}
Suppose that $\A$ is a deterministic matrix, $\sigma_{\min}$ denotes the smallest positive singular value
of $\A_{\Omega}$ and $\x$ is a $K$-sparse signal. For any fixed $0\leq k\leq |\Omega|-1$,
suppose that $\mathcal{S}_k\subseteq \Omega$, $|\mathcal{S}_k|=k$ and
the inequality in \eqref{e:csk} holds with $\mathcal{S}=\Omega\setminus\mathcal{S}_k$
for certain nondecreasing function $\phi(t)$.
Then $\mathcal{S}_{k+1}\subseteq \Omega$ and $|\mathcal{S}_{k+1}|=k+1$  provided that
\beq
\label{e:main}
\|\A_{\Omega^c}^\top\u_k\|_{\infty}<\frac{\sigma_{\min}}{\sqrt{\phi(K-k)}},
\eeq
where
\beq
\label{e:uk}
\u_k=\frac{\P^{\perp}_{\mathcal{S}_k}\A_{\Omega\setminus \mathcal{S}_k}\x_{\Omega\setminus \mathcal{S}_k}}
{\|\P^{\perp}_{\mathcal{S}_k}\A_{\Omega\setminus \mathcal{S}_k}\x_{\Omega\setminus \mathcal{S}_k}\|_{2}}.
\eeq
\end{lemma}

\begin{IEEEproof}
See Appendix \ref{ss:pflmain}.
\end{IEEEproof}

We next introduce Lemma \ref{l:Gaussian1}  which essentially gives a lower bound on the probability of
the left-hand side of \eqref{e:main} being less than a given constant $\epsilon$
(by setting $\A_{\Omega^c}^\top=\B$ and $\ell=1$).

\begin{lemma}
\label{l:Gaussian1}
Suppose that $\B\in \mathbb{R}^{m\times p}$ is a random matrix, whose entries independently
and identically follow the Gaussian distribution $\mathcal{N}(0, 1/m)$,
and $\ell$ is an arbitrary positive integer.
Let $\u_i\in \mathbb{R}^{m}$, which is independent with $\B$, satisfy $\|\u_i\|_2\leq 1$
for $1\leq i\leq \ell$.
Then, for any positive $\epsilon_i, 1\leq i\leq \ell$, we have
\beq
\label{e:pbd}
\mathbb{P}(\bigcap_{i=1}^\ell(\|\B^{\top}\u_i\|_{\infty}\leq\epsilon_i))\geq
\prod_{i=1}^\ell
\left(1-\frac{e^{-\epsilon_i^2m/2}}{\sqrt{\pi m/2}\epsilon_i}\right)^{p}.
\eeq
\end{lemma}

\begin{IEEEproof}
See Appendix \ref{ss:pflGauss}.
\end{IEEEproof}

Note that although there are some connections between Lemma \ref{l:Gaussian1}
and \cite[Propostion 4]{TroG07b}, there are two main differences between them.
Firstly, the column vector $\z$ in \cite[Propostion 4]{TroG07b}
has been extended to a matrix $\B$ in Lemma \ref{l:Gaussian1}.
Secondly, Lemma \ref{l:Gaussian1} is sharper than
\cite[Propostion 4]{TroG07b}
when $\epsilon>1/\sqrt{\pi m/2}$ if we assume $\B$ is a column vector.

To show Theorem \ref{t:noiseless}, we also need to introduce the following lemma from \cite{DavS02}.
This lemma together with Lemma \ref{l:Gaussian1} characterize the probability of \eqref{e:main} holds.

\begin{lemma}
\label{l:sigmaK}
Suppose that $\B\in \mathbb{R}^{m\times p}$ is a random matrix, whose entries independently
and identically follow the Gaussian distribution $\mathcal{N}(0, 1/m)$.
If $m\geq p$, then the smallest singular value $\sigma_{\min}$ of $\B$ satisfies
\begin{align}
\label{e:sigmaK}
\mathbb{P}(\sigma_{\min}\geq 1-\sqrt{p/m}-\epsilon)\geq1-e^{-\epsilon^2m/2}
\end{align}
for any given $\epsilon>0$.
\end{lemma}

To prove Theorem \ref{t:nbofob}, the following Lemma is also needed.

\begin{lemma}
\label{l:varphi}
Let $\varphi(t)=\sqrt{t}e^{-1/t},\,\; 0<t<2$.
Suppose that $0<t_1\leq t_2\leq \ldots\leq t_p<2$ for some integer $p$, then
\begin{align}
\label{e:varphi2}
\sum_{i=1}^p\varphi(t_i)\leq  \frac{\sum_{i=1}^pt_i}{\sqrt{t_p}e^{1/t_p}}.
\end{align}
\end{lemma}

\begin{IEEEproof}
See Appendix \ref{ss:varphi}.
\end{IEEEproof}

\subsection{Proof of Theorem \ref{t:noiseless}}
\label{ss:pb}

In the following, we use Lemmas \ref{l:main}--\ref{l:sigmaK} to prove Theorem \ref{t:noiseless}.
\begin{IEEEproof}
Without loss of generality, we assume $\x$ has exact $K$ nonzero entries.
Then, to show the theorem, it suffices to show that $\mathcal{S}_k\subseteq \Omega$
and $|\mathcal{S}_k|=k$ for $0\leq k\leq K-1$.
Hence, by Lemma \ref{l:main} and induction, it suffices to show that
\eqref{e:main} hold for $0\leq k\leq K-1$.

To simplify notation, we denote event $F_k$ as
\begin{align}
F_k:=\left\{\|\A_{\Omega^c}^\top\u_k\|_{\infty}<\frac{\sigma_{\min}}{\sqrt{\phi(K-k)}}
\right\}, \,\;0\leq k\leq K-1,
\label{e:Fk}
\end{align}
where $\sigma_{\min}$ denotes the smallest singular value of $\A_{\Omega}$.
Then by the above analysis, \eqref{e:E} and Lemma \ref{l:main}, we have
\[
\mathbb{P}(\mathbb{S})\geq\mathbb{P}(\bigcap_{k=0}^{K-1}F_k).
\]
For any
\[
0<\epsilon< 1-\sqrt{\frac{K}{m}}-\sqrt{\frac{2\phi(K)}{m\pi}},
\]
by \eqref{e:t},
\[
\eta=1-\sqrt{\frac{K}{m}}-\epsilon> \sqrt{\frac{2\phi(K)}{m\pi}}.
\]
Hence,
\beq
1>\frac{e^{-\frac{\eta^2m}{2\phi(k)}}}{\sqrt{\frac{\pi m}{2\phi(k)}}\eta},\,\; 1\leq k\leq K.
\label{e:positive}
\eeq
Furthermore, we have
\begin{align}
\label{e:pbproof}
\mathbb{P}(\mathbb{S})\geq&\mathbb{P}(\bigcap_{k=0}^{K-1}F_k)
\geq\mathbb{P}(\bigcap_{k=0}^{K-1}F_k, \sigma_{\min}\geq \eta)\nonumber\\
=&\mathbb{P}(\bigcap_{k=0}^{K-1}F_k|\sigma_{\min}\geq \eta)\mathbb{P}(\sigma_{\min}\geq \eta)\nonumber\\
\overset{(a)}{\geq}&\mathbb{P}\left(\bigcap_{k=0}^{K-1}\left(\|\A_{\Omega^c}^\top\u_k\|_{\infty}
<\frac{\eta}{\sqrt{\phi(K-k)}}\right)\right)\nonumber\\
\times&\mathbb{P}(\sigma_{\min}\geq \eta)\nonumber\\
\overset{(b)}{\geq}&\prod_{k=0}^{K-1}\left(1-\frac{e^{-\frac{\eta^2m}{2\phi(K-k)}}}
{\sqrt{\frac{\pi m}{2\phi(K-k)}}\eta}\right)^{(n-K)}(1-e^{-\frac{\epsilon^2m}{2}})\nonumber\\
=&(1-e^{-\frac{\epsilon^2m}{2}})
\prod_{k=1}^{K}\left(1-\frac{e^{-\frac{\eta^2m}{2\phi(k)}}}
{\sqrt{\frac{\pi m}{2\phi(k)}}\eta}\right)^{(n-K)},
\end{align}
where (a) is from \eqref{e:Fk}
and (b) follows from
\eqref{e:t}, Lemmas \ref{l:Gaussian1} and \ref{l:sigmaK}, \eqref{e:positive}
and the fact that
\begin{align*}
&\mathbb{P}\left(\bigcap_{k=0}^{K-1}\left(\|\A_{\Omega^c}^\top\u_k\|_{\infty}
<\frac{\eta}{\sqrt{\phi(K-k)}}\right)\right)\\
=&\mathbb{P}\left(\bigcap_{k=0}^{K-1}\left(\|\A_{\Omega^c}^\top\u_k\|_{\infty}
\leq\frac{\eta}{\sqrt{\phi(K-k)}}\right)\right).
\end{align*}
Hence, Theorem \ref{t:noiseless} holds.
\end{IEEEproof}

In the following, we explain the connections and differences between the
proofs of \cite[Theorem 6]{TroG07b}  and Theorem \ref{t:noiseless}.
Same as the proof of \cite[Theorem 6]{TroG07b},
to lower bound $\mathbb{P}(\mathbb{S})$,
we lower bound the probability that
the OMP algorithm can find an index in $\Omega$ in each iteration
under the condition that $\sigma_{\min}$ is not smaller than a constant
(the constant is $\eta$ in the proof of Theorem \ref{t:noiseless});
for more details, see the first two inequalities of \eqref{e:pbproof}.
Compared with the proof of \cite[Theorem 6]{TroG07b},
we give a sharper lower bound
by utilizing Lemmas \ref{l:main} and \ref{l:Gaussian1} which
use the sparsity property of $\x$, \eqref{e:csk},
some techniques from matrix theory and
a sharper upper bound on the Gaussian Q-function
(for more details, see the proofs of Lemmas \ref{l:main} and \ref{l:Gaussian1}).

\subsection{Proof of Theorem \ref{t:nbofob}}
\label{ss:pfnm}

In the following, we use Theorem \ref{t:noiseless} and Lemma \ref{l:varphi} to prove Theorem \ref{t:nbofob}.

\begin{IEEEproof}
Let
\begin{align}
\label{e:epsilon0}
\epsilon_0&=\sqrt{\frac{2\ln(n/\delta)}{m}},\,\;
\eta_0=1-\sqrt{\frac{K}{m}}-\epsilon_0,\\
P(\epsilon_0)&=(1-e^{-\frac{\epsilon_0^2m}{2}})\prod_{k=1}^{K}
\left(1-\frac{e^{-\frac{\eta_0^2m}{2\phi(k)}}}{\sqrt{\frac{\pi m}{2\phi(k)}}\eta_0}\right)^{(n-K)}.
\label{e:pfunc}
\end{align}
Then, by Theorem \ref{t:noiseless}, to show \eqref{e:prob2}, it suffices to show that
\beq
\label{e:valideps}
\epsilon_0\leq1-\sqrt{\frac{K}{m}}-\sqrt{\frac{2\phi(K)}{m\pi}}
\eeq
and
\begin{align}
\label{e:pfunclb}
\mathbb{P}(\epsilon_0)\geq 1-\left(\frac{\delta}{n}+\frac{\delta^{\beta}}{\sqrt{\pi\beta}}\right).
\end{align}
By \eqref{e:m}, we have
\begin{align*}
&\left(\sqrt{\frac{2\ln(n/\delta)}{m}}+\sqrt{\frac{K}{m}}+\sqrt{\frac{2\phi(K)}{m\pi}}\right)^2\\
=&\frac{1}{m}\left(\sqrt{2\ln(n/\delta)}+\sqrt{K}+\sqrt{\frac{2\phi(K)}{\pi}}\right)^2<1,
\end{align*}
thus by \eqref{e:epsilon0}, \eqref{e:valideps} holds.

In the following, we prove \eqref{e:pfunclb}.
By induction, one can easily show that
\[
\prod_{k=1}^{K}(1-a_it)\geq 1-(\sum_{k=1}^{K}a_i)t, \,\; a_i\geq 0, t\geq0.
\]
Hence, by \eqref{e:pfunc}, we have
\begin{align}
\label{e:pfunclb2}
\mathbb{P}(\epsilon_0)\geq&(1-e^{-\frac{\epsilon_0^2m}{2}})
\sum_{k=1}^{K}
\left(1-\frac{n-K}{\sqrt{\pi}}\sqrt{\frac{2\phi(k)}{\eta_0^2m}}e^{-\frac{\eta_0^2m}{2\phi(k)}}\right)\nonumber\\
\geq&1-e^{-\frac{\epsilon_0^2m}{2}}-\frac{n-K}{\sqrt{\pi}}\sum_{k=1}^{K}
\left(\sqrt{\frac{2\phi(k)}{\eta_0^2m}}e^{-\frac{\eta_0^2m}{2\phi(k)}}\right)\nonumber\\
\geq&1-e^{-\frac{\epsilon_0^2m}{2}}-\frac{n-K}{\sqrt{\pi}}\sqrt{\frac{\eta_0^2m}{2\phi(K)}}
e^{-\frac{\eta_0^2m}{2\phi(K)}}\frac{2\sum_{k=1}^{K}\phi(k)}{\eta_0^2m},
\end{align}
where the last inequality is from Lemma \ref{l:varphi}  with $t_i=\frac{2\phi(i)}{\eta_0^2m}$,
which are less than 2 for $1\leq i\leq K$ according to \eqref{e:tub} below.

In the following, we give upper bounds on the last  two terms of the right-hand side
of \eqref{e:pfunclb2}. By \eqref{e:epsilon0}, we have
\begin{align}
\label{e:pfuncub1}
e^{-\frac{\epsilon_0^2m}{2}}=e^{-\ln(n/\delta)}=\frac{\delta}{n}.
\end{align}
To give an upper bound on the last term of the right-hand side of \eqref{e:pfunclb2},
we first given a lower bound on $\sqrt{\frac{\eta_0^2 m}{2\phi(K)}}$. By \eqref{e:epsilon0}, we have
\begin{align}
\label{e:tub}
\sqrt{\frac{\eta_0^2 m}{2\phi(K)}}
=&\left(1-\sqrt{\frac{K}{m}}-\sqrt{\frac{2\ln(n/\delta)}{m}}\right)\sqrt{\frac{ m}{2\phi(K)}}\nonumber\\
=&\sqrt{\frac{ m}{2\phi(K)}}-\sqrt{\frac{K}{2\phi(K)}}-\sqrt{\frac{\ln(n/\delta)}{\phi(K)}}
\nonumber\\
\geq&\sqrt{\beta\ln(n/\delta)},
\end{align}
where the last inequality follows from \eqref{e:m}. Thus,
\beq
\label{e:efub}
e^{-\frac{\eta_0^2m}{2\phi(K)}}\leq e^{-\beta\ln(n/\delta)}=(\frac{\delta}{n})^{\beta}.
\eeq
Therefore, we have
\begin{align}
\label{e:lastterm}
&\frac{n-K}{\sqrt{\pi}}\sqrt{\frac{\eta_0^2m}{2\phi(K)}}
e^{-\frac{\eta_0^2m}{2\phi(K)}}\frac{2\sum_{k=1}^{K}\phi(k)}{\eta_0^2m}\nonumber \\
=&\frac{n-K}{\sqrt{\pi}}\sqrt{\frac{2\phi(K)}{\eta_0^2m}}
e^{-\frac{\eta_0^2m}{2\phi(K)}}\frac{\sum_{k=1}^{K}\phi(k)}{\phi(K)}\nonumber \\
\leq&\frac{n-K}{\sqrt{\pi}}\frac{1}{\sqrt{\beta\ln(n/\delta)}}\frac{\delta^{\beta}}{n^{\beta}}
\frac{\sum_{k=1}^{K}\phi(k)}{\phi(K)}
\leq \frac{\delta^{\beta}}{\sqrt{\pi\beta}},
\end{align}
where the first inequality is from \eqref{e:tub} and \eqref{e:efub},
and the second inequality is from \eqref{e:alpha}.
Then by the above inequality, \eqref{e:pfunclb2}, \eqref{e:pfuncub1} and \eqref{e:lastterm},
we can see that \eqref{e:pfunclb} holds, and hence the theorem holds.
\end{IEEEproof}

\section{Simulation tests}
\label{s:sim}

In this section, we perform simulation tests to illustrate our main results presented in
Section \ref{s:main} and compare them with existing ones.

\subsection{Simulation tests for the probability of exact recovery}
\label{ss:sim1}

\begin{figure}[t]
\centering
\includegraphics[width=3.2in]{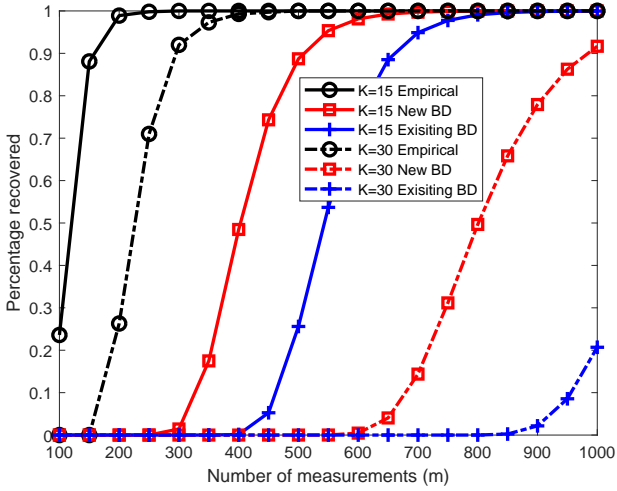}
\caption{Empirical Probability and lower bounds on recovering $K$-sparse flat signals in dimension $n=1024$ over 1000 realizations}
\label{f:prob1}
\end{figure}

\begin{figure}[t]
\centering
\includegraphics[width=3.2in]{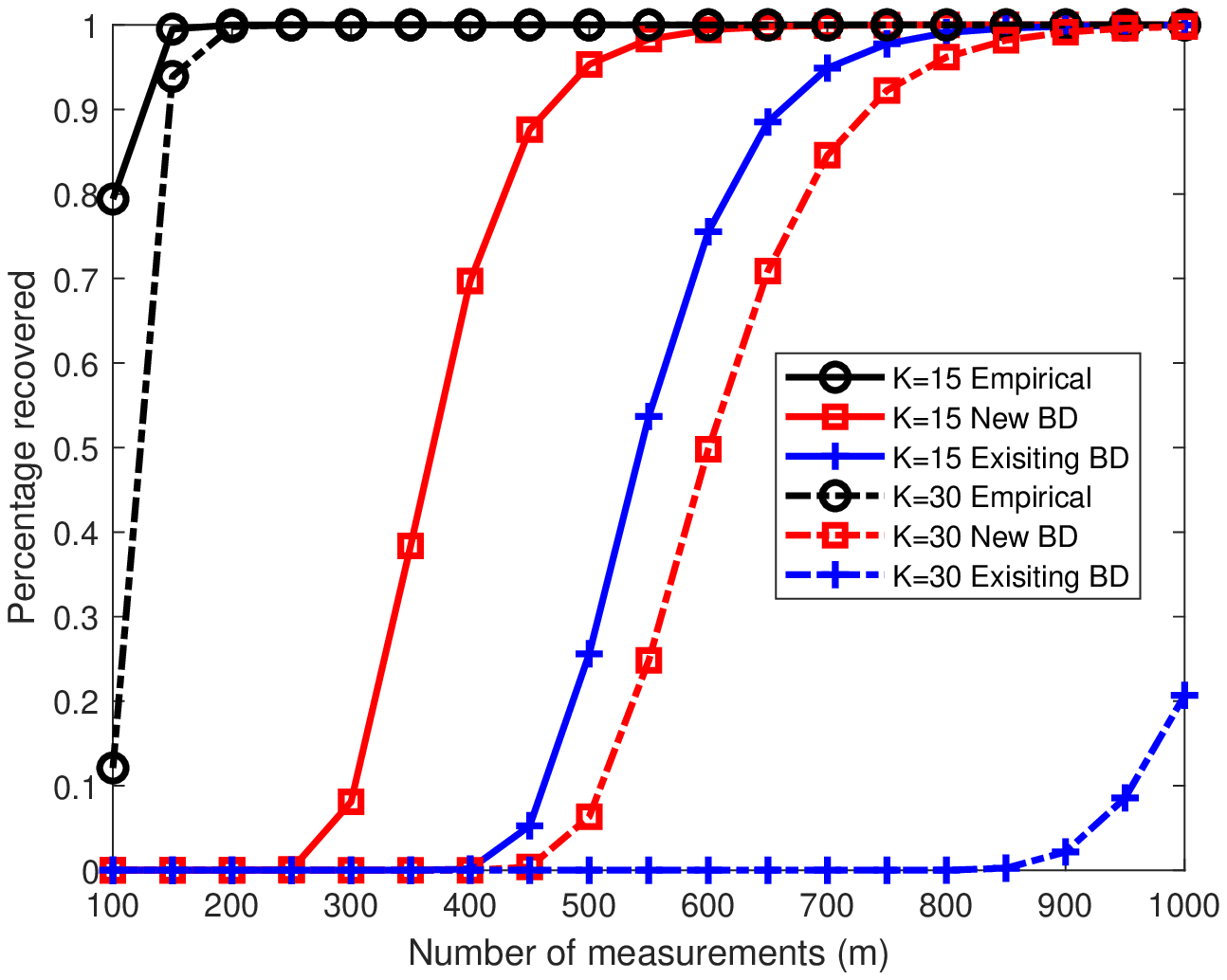}
\caption{Empirical Probability and lower bounds on recovering $K$-sparse 1.1-strongly-decaying signals
in dimension $n=1024$  over 1000 realizations}
\label{f:prob21}
\end{figure}

\begin{figure}[t]
\centering
\includegraphics[width=3.2in]{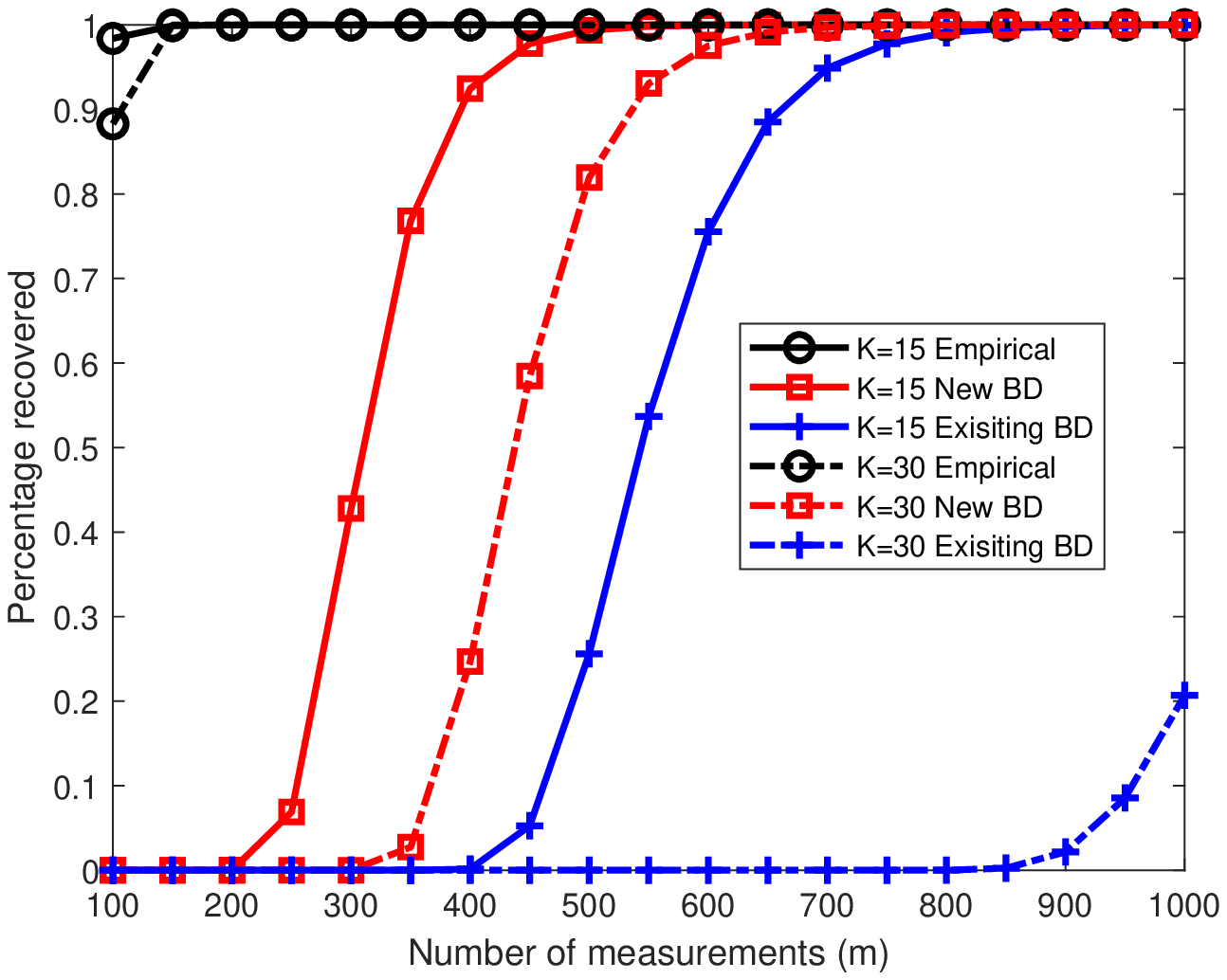}
\caption{Empirical Probability and lower bounds on recovering $K$-sparse 1.2-strongly-decaying signals
 in dimension $n=1024$  over 1000 realizations}
\label{f:prob22}
\end{figure}

\begin{figure}[t]
\centering
\includegraphics[width=3.2in]{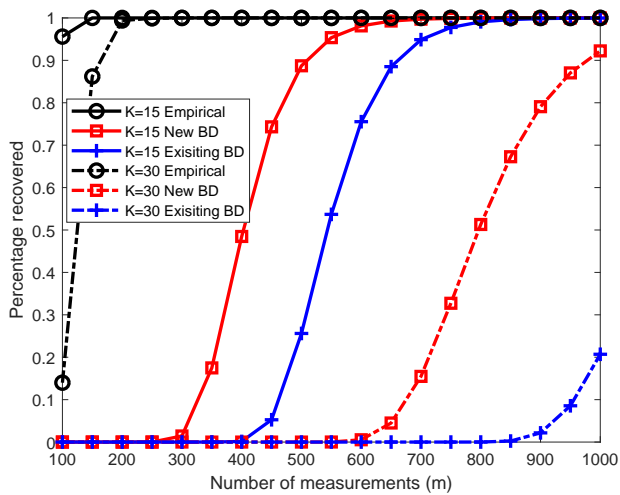}
\caption{Empirical Probability and lower bounds on recovering $K$-sparse Gaussian signals in dimension $n=1024$  over 1000 realizations}
\label{f:prob3}
\end{figure}

In this subsection, we conduct simulation tests to illustrate Theorem \ref{t:noiseless},
Corollaries \ref{c:regular}--\ref{c:Gauss} and compare them with \cite[Theorem 6]{TroG07b}.

We generated 1000 realizations of linear model \eqref{e:model}.
More specifically, for each fixed $m$, $n$ and $K$, and for each realization,
we generated a matrix $\A\in \mathbb{R}^{m\times n}$ with i.i.d. $\mathcal{N}(0, 1/m)$ entries;
we randomly selected $K$ elements from the set $\{1,2,\ldots, n\}$ to form the support $\Omega$ of $\x$,
and then generated an $\x\in \mathbb{R}^{n}$ according to the following four cases:
\begin{enumerate}
\item $x_i=1$ for $i\in \Omega$ ;
\item The $i$-th element of $\x_{\Omega}$ is $1.1^{K-i}$ for $i\in \Omega$;
\item The $i$-th element of $\x_{\Omega}$ is $1.2^{K-i}$ for $i\in \Omega$;
\item $\x_{\Omega}=\mbox{randn}(K,1)$, where randn is a MATLAB built-in function.
\end{enumerate}

After generating $\A$ and $\x$, we set $\y=\A\x$.
Hence, for each fixed $m$, $n$, $K$ and for each case,
we have $1000$ linear models in the form of \eqref{e:model}.
Then, we use OMP (i.e., Algorithm \ref{a:OMP}) to reconstruct $\x$,
and count the number of exactly recovery of $\x$ (note that $\x$ is thought as exactly recovered if
the 2-norm of the difference between the returned $\x$ and generated $\x$ is not larger than $10^{-10}$).
Finally, we divide the number of exactly recovery of $\x$ by 1000 and denote it as ``Empirical".

We compute the right-hand side of \eqref{e:prob} with $\phi(t)=t$ and $\phi(t)$ being defined
by \eqref{e:phit2} for Cases 1 and 4, respectively.
Since $\x$ from Cases 2 and 3 are $K$-sparse $\alpha$-strongly-decaying signals with
$\alpha=1.1$ and $\alpha=1.2$, respectively, we compute the right-hand side of \eqref{e:prob}
with $\phi(t)$ being defined by \eqref{e:phit}  with $\alpha=1.1$ and $\alpha=1.2$ for Cases 2 and 3,
respectively. All of these values are denoted as ``New BD".
To compare Corollaries \ref{c:regular}--\ref{c:Gauss} with \cite[Theorem 6]{TroG07b},
we also compute the right-hand side of \eqref{e:probtrop} and denote it as ``Existing BD".
Since the lower bound on $\mathbb{P}(\mathbb{S})$ given by \cite[Theorem 6]{TroG07b}  uses the sparsity property of $\x$ only,
``Existing BD" are the same for all the four cases.

Figs. \ref{f:prob1}-\ref{f:prob3} respectively display ``Empirical",
``New BD" and ``Existing BD" for $m=100:50:1000$ and $n=1024$ with $K=15$ and $K=30$
for $\x$ from Cases 1-4.
Note that from Corollary \ref{l:Gaussian},  ``New BD" holds with probability $\nu$
satisfies \eqref{e:pGaussian} for Case 4.

Figs. \ref{f:prob1}-\ref{f:prob3} show that ``New BD" are much tighter than ``Existing BD"
for all the four cases which indicates that the lower bounds on $\mathbb{P}(\mathbb{S})$ given by
Corollaries \ref{c:regular}--\ref{c:Gauss} are much sharper than that given by \cite[Theorem 6]{TroG07b}.
They also show that OMP  has significantly better recovery performance in recovering
$\alpha$-strongly-decaying and Gaussian sparse signals than recovering flat sparse signals.

The black lines in Figs. \ref{f:prob21}-\ref{f:prob22} show that the recovery performance of the OMP algorithm
for recovering $\alpha$-strongly-decaying sparse signals becomes better as $\alpha$ gets larger.

Figs. \ref{f:prob1}-\ref{f:prob3} also show that the gap between the new lower bound on $\mathbb{P}(\mathbb{S})$
and the empirical $\mathbb{P}(\mathbb{S})$ is very large, which indicates that the new lower bounds given by
Theorem \ref{t:noiseless} and Corollaries \ref{c:regular}--\ref{c:Gauss} are not sharp and there is much room to improve them.
However, this may be difficult. Before giving reasons to explain the difficulties,
we describe some reasons leading to the loose bound.
\begin{itemize}
\item Although we give closed form expressions for $\phi(t)$ by using some prior information of $\x$,
the difference between the right-hand side and left-hand side of \eqref{e:csk}
may be large. This leads to the gap between the two sides of \eqref{e:linflbd} being large,
as a result, the gap between the two sides of \eqref{e:main} is also large.
By the proof of Theorem \ref{t:noiseless}, this causes the bound given by \eqref{e:prob} being loose.
\item The difference between the two sides of \eqref{e:sigmaK} may also
large, which causes that the bound on $\mathbb{P}(\mathbb{S})$ being loose.
\item It is also possible that the second inequality in \eqref{e:pbproof} is not sharp,
which causes the bound on $\mathbb{P}(\mathbb{S})$ being not sharp.
\end{itemize}

To improve the bound on $\mathbb{P}(\mathbb{S})$, at least one of the above drawbacks need to be
addressed, so it may be difficult to improve the bound.

\begin{figure}[t]
\centering
\includegraphics[width=3.2in]{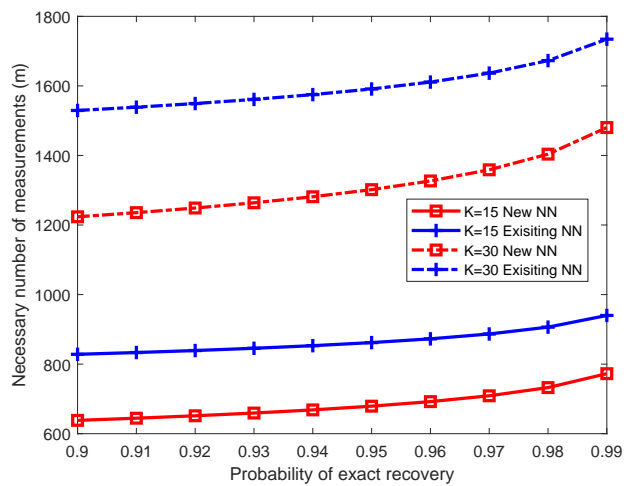}
\caption{Lower bounds on the necessary number of measurements $m$ for recovering $K$-sparse flat signals in dimension $n=1024$}
\label{f:m1}
\end{figure}

\begin{figure}[t]
\centering
\includegraphics[width=3.2in]{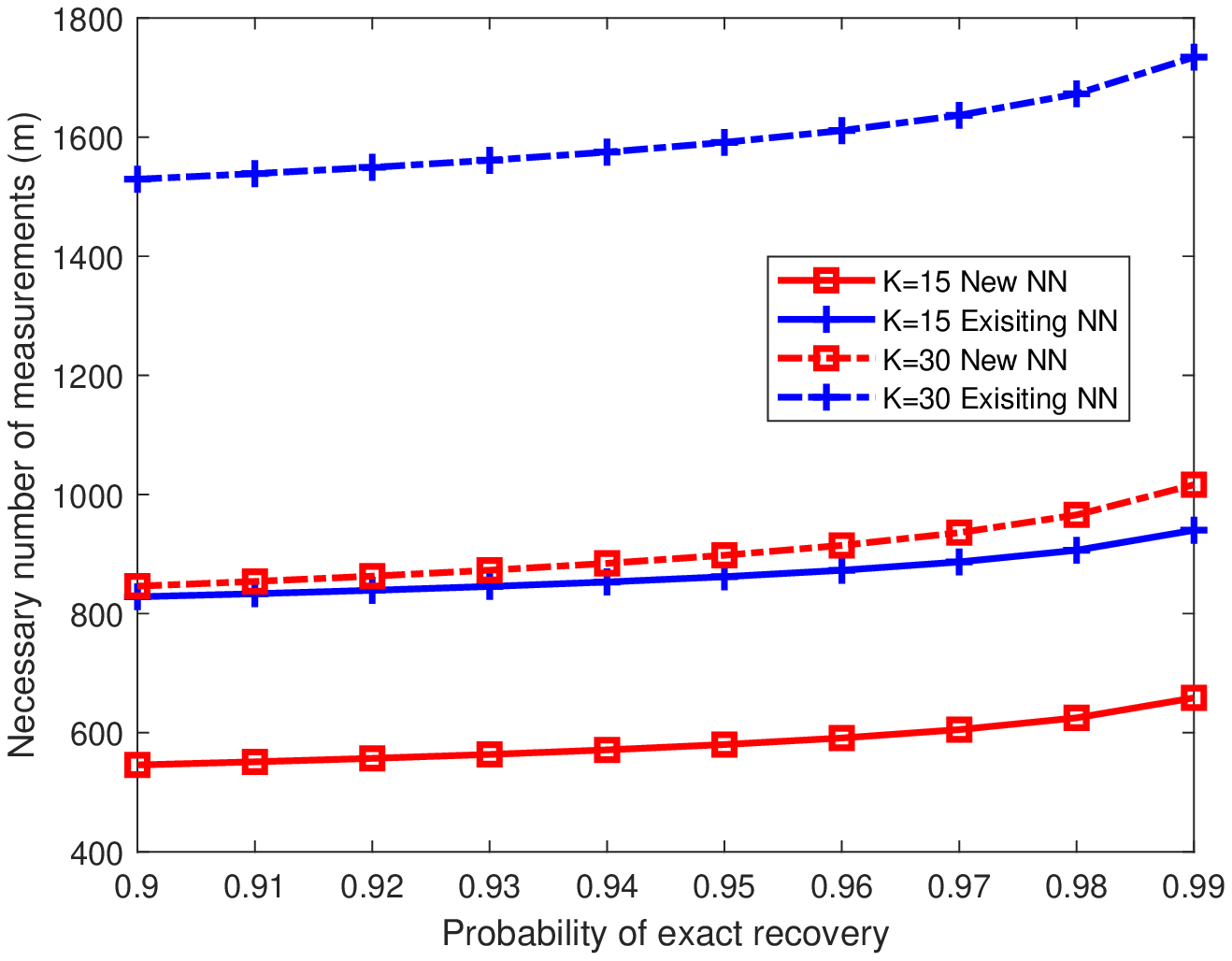}
\caption{Lower bounds on the necessary number of measurements $m$ for recovering $K$-sparse 1.1-strongly-decaying signals in dimension $n=1024$}
\label{f:m21}
\end{figure}

\begin{figure}[t]
\centering
\includegraphics[width=3.2in]{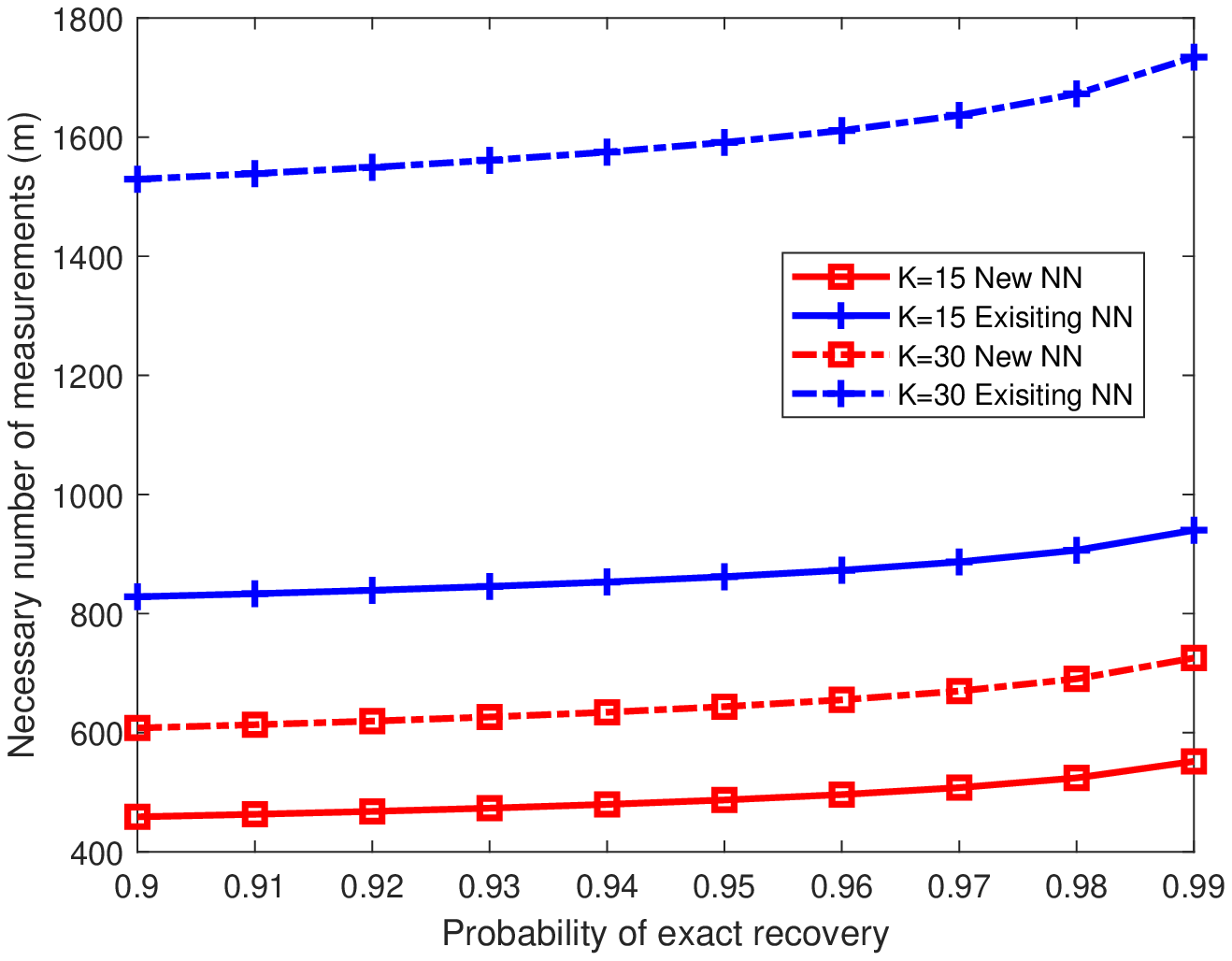}
\caption{Lower bounds on the necessary number of measurements $m$ for recovering $K$-sparse 1.2-strongly-decaying signals  in dimension $n=1024$}
\label{f:m22}
\end{figure}

\begin{figure}[t]
\centering
\includegraphics[width=3.2in]{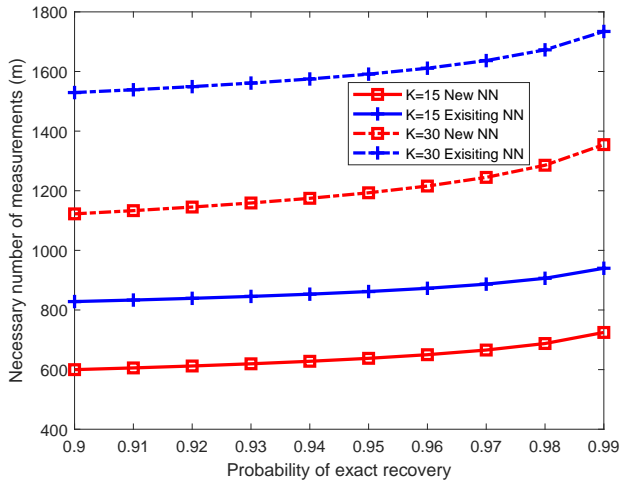}
\caption{Lower bounds on the necessary number of measurements $m$ for recovering $K$-sparse Gaussian signals in dimension $n=1024$}
\label{f:m3}
\end{figure}

\subsection{Simulation tests for necessary number of measurements}
\label{ss:sim2}

In this subsection, we perform simulations to illustrate Theorem \ref{t:nbofob},
Corollaries \ref{c:nbofobcoro}--\ref{c:nbofob4} and compare them with \cite[Corollary 7]{TroG07b}.

We perform simulations to test the necessary number of measurements $m$ to guarantee that
$\mathbb{P}(\mathbb{S})$ is not smaller than $0.90:0.01:0.99$ for recovering the four classes of $K$-sparse signals
defined in Section \ref{ss:sim1}, so we set $\zeta=0.1:-0.01:0.01$ (see \eqref{e:prob3}).

For each fixed $\zeta$, by setting $\delta=\frac{n\sqrt{\pi}\zeta}{n+\sqrt{\pi}}$,
we respectively compute the right-hand side of \eqref{e:m} with
$\phi(t)=t$ and with $\phi(t)$ being defined by \eqref{e:phit2} for $\x$ from Cases 1 and 4, respectively.
Since $\x$ from Cases 2 and 3 are $K$-sparse $\alpha$-strongly-decaying signals with
$\alpha=1.1$ and $\alpha=1.2$, respectively, by setting $\delta=\frac{n\sqrt{\pi}\zeta}{n+\sqrt{\pi}}$,
we respectively compute the right-hand side of \eqref{e:m}
with $\phi(t)$ being defined by \eqref{e:phit}  with $\alpha=1.1$ and $\alpha=1.2$ for Cases 2 and 3.
All of these values are denoted by ``New NN".
To compare Theorem \ref{t:nbofob},
Corollaries \ref{c:nbofob}--\ref{c:nbofob4} with \cite[Corollary 7]{TroG07b},
we also compute the right-hand side of \eqref{e:mtrop} with $\delta$ being defined in \eqref{e:deltatrop}
and denote it by ``Existing NN".
Since the lower bound on $\mathbb{P}(\mathbb{S})$ given by \cite[Theorem 6]{TroG07b}  uses the sparsity property
of $\x$ only, ``Existing NN" are the same for the four cases.

Figs. \ref{f:m1}-\ref{f:m3} respectively show ``New NN" and ``Existing NN"
for $\zeta=0.1:-0.01:0.01$ (which ensures that $\mathbb{P}(\mathbb{S})$ is not smaller than $0.90:0.01:0.99$)
and $n=1024$ with $K=15$ and $K=30$ for recovering $\x$ from Cases 1-4 which are defined in Section \ref{ss:sim1}.
Note that from Lemma \ref{l:Gaussian}, ``New NN" guarantees that $\mathbb{P}(\mathbb{S})$ is not smaller than $0.90:0.01:0.99$ with probability $\nu$
satisfies \eqref{e:pGaussian} for Case 4.

Figs. \ref{f:m1}-\ref{f:m3} show that ``New NN" are much smaller than ``Existing NN"
for all the four cases which indicates that Theorem \ref{t:nbofob},
Corollaries \ref{c:nbofob}--\ref{c:nbofob4} are much better than \cite[Corollary 7]{TroG07b}
in characterizing the necessary number of measurements $m$ which ensures that
$\mathbb{P}(\mathbb{S})$ is not smaller than a given probability.
They also show that to guarantee $\mathbb{P}(\mathbb{S})$ is not lower than a target probability,
many more measurements are needed for recovering flat signals than those required by recovering
$\alpha$-strongly-decaying sparse signals and Gaussian sparse signals especially when $K$ is large.

Figs. \ref{f:m21}-\ref{f:m22} show that to guarantee a target recovery performance of OMP
for recovering $\alpha$-strongly-decaying signals,
the necessary number of measurements $m$ decreases as $\alpha$ increases.

Note that, in the tests, we also found that the empirical necessary number of measurements $m$
is much smaller than the new bound.
This is because the gap between the new theoretical bound on $\mathbb{P}(\mathbb{S})$
and the empirical $\mathbb{P}(\mathbb{S})$ is large.
\subsection{Simulation tests for the effect of $\frac{\|\x\|_1^2}{\|\x\|_2^2}$ on the probability of exact recovery}
\label{ss:sim3}

It is interesting to investigate how $\phi(t)$ affects the probability of exact recovery of $K$-sparse
signals $\x$ which satisfies \eqref{e:csk} with $K$-iterations of OMP,
but since it is difficult to find $\phi(t)$, we perform simulations to show how
$\frac{\|\x\|_1^2}{\|\x\|_2^2}$ affects the probability of exact recovery.

We choose $n=1024$ and $K=15, 30$.
To ensure the probability of exact recovery is not too low, for $K=15$, we take $m=60,80,100$, while for $K=30$, we take $m=120,140,160$.
We did the tests over 10000 runs.
For each run and for each fixed $K$ and $m$, we randomly generated a matrix $\A$ with i.i.d. $\mathcal{N}(0, 1/m)$ entries,
and 6 $\x$'s whose supports $\Omega$ have exact $K$ elements and are randomly chosen from $\{1,2,\ldots, 1024\}$.
Theses $\x$'s were generated according to:
\begin{enumerate}
\item $x_i=1$ for $i\in \Omega$;
\item $x_i$ independently and identically follow the uniform distribution on $[-\sqrt{3},\sqrt{3}]$ for $i\in \Omega$;
\item $\x_{\Omega}=\mbox{randn}(K,1)$, where randn is a MATLAB built-in function;
\item The $i$-th element of $\x_{\Omega}$ is $1.2^{K-i}$ for $i\in \Omega$;
\item $x_i$ independently and identically follow the exponential distribution with $\lambda=1$ for $i\in \Omega$;
\item $x_i$ independently and identically follow the Poisson distribution with $\lambda=1$ for $i\in \Omega$.
\end{enumerate}
Note that the variances of $x_i$ from Cases 2,3,5 and 6 are 1 for $i\in \Omega$.

After generating $\A$ and $\x$, we set $\y=\A\x$, use Algorithm \ref{a:OMP} to reconstruct $\x$,
calculate the empirical probability of exact recovery (see Section \ref{ss:sim1}),
and denote it as $\mathbb{P}(\mathbb{S})$.
We also compute the average $\frac{\|\x\|_1^2}{\|\x\|_2^2}$.

\begin{table}[t]
\caption{Recovery of $15$-sparse signals in dimension $n=1024$ over 10000 realizations}
\vspace*{-2mm}
\centering
\begin{tabular}{||c||c|c||c|c||c|c||}
 \hline
& \multicolumn{2}{c||}{m=60} & \multicolumn{2}{c||}{m=80} & \multicolumn{2}{c||}{m=100} \\ \hline
& $\frac{\|\x\|_1^2}{\|\x\|_2^2}$  &  $\mathbb{P}(\mathbb{S})$ &   $\frac{\|\x\|_1^2}{\|\x\|_2^2}$ &  $\mathbb{P}(\mathbb{S})$  &   $\frac{\|\x\|_1^2}{\|\x\|_2^2}$  &  $\mathbb{P}(\mathbb{S})$ \\ \hline
Case 1& 15 &0&15 &0.021&15 &0.225\\ \hline
Case 2& 11.4458 &0.221&11.3587 &0.527 &11.4035 &0.771\\ \hline
Case 3& 9.8326 &0.256&9.8819 &0.790&9.8969 &0.965\\ \hline
Case 4& 9.6591 &0.272&9.6591 &0.871&9.6591 &0.987\\ \hline
Case 5& 8.2491 &0.400&8.3411  &0.892&8.3028 &0.978\\ \hline
Case 6& 7.9221 &0.640&7.7313  &0.946&7.8983 &0.981\\ \hline
\end{tabular}
\label{tb:1}
\end{table}

\begin{table}[t]
\caption{Recovery of $30$-sparse signals in dimension $n=1024$ over 10000 realizations}
\vspace*{-2mm}
\centering
\begin{tabular}{||c||c|c||c|c||c|c||}
 \hline
& \multicolumn{2}{c||}{m=120} & \multicolumn{2}{c||}{m=140} & \multicolumn{2}{c||}{m=160} \\ \hline
&$\frac{\|\x\|_1^2}{\|\x\|_2^2}$  &  $\mathbb{P}(\mathbb{S})$ &   $\frac{\|\x\|_1^2}{\|\x\|_2^2}$  &  $\mathbb{P}(\mathbb{S})$  &   $\frac{\|\x\|_1^2}{\|\x\|_2^2}$  &  $\mathbb{P}(\mathbb{S})$ \\ \hline
Case 1& 30 &0&30 &0.002&30 &0.018\\ \hline
Case 2& 22.6196 &0.221&22.5235 &0.527 &22.5841 &0.771\\ \hline
Case 3& 19.4146 &0.480&19.4692 &0.758&19.4546 &0.907 \\ \hline
Case 4& 10.9077 &0.985&10.9077 &1&10.9077 &1\\ \hline
Case 5& 15.8608 &0.687&15.8855 &0.906&15.7934 &0.972\\ \hline
Case 6& 15.3644 &0.951&15.3518&0.995&15.3607 &0.999\\ \hline
\end{tabular}
\label{tb:2}
\end{table}

Tables \ref{tb:1} and \ref{tb:2} respective show the average $\frac{\|\x\|_1^2}{\|\x\|_2^2}$
and $\mathbb{P}(\mathbb{S})$ for $K=15$ and $K=30$ for $\x$ from the above six cases.
From these two tables, we can see that
$\mathbb{P}(\mathbb{S})$ tends to increase as $\frac{\|\x\|_1^2}{\|\x\|_2^2}$
decreases for fixed $K$ and $m$, and
$\mathbb{P}(\mathbb{S})$ becomes larger as $m$ becomes larger for fixed $n$ and $K$.
They also show that, among the six cases, OMP has the best and worst recoverability for recovering
$\x$ whose nonzero entries follow the Poisson distribution and are the same, respectively.

\section{Conclusion and Future work}
\label{s:con}

In this paper, we developed lower bounds on the probability of exact recovery
of $K$-sparse signals $\x$ using $K$ iterations of OMP  and
the necessary number of measurements
which guarantees that $\x$  can be exactly recovered with OMP in $K$ iterations
with overwhelming probability.
These lower bounds depend on a function $\phi(t)$ which is used to measure
the variations in the magnitudes of the nonzero entries of $\x$.
By exploring the prior information of the nonzero entries of $\x$,
we developed  closed-form expressions of $\phi(t)$ for $K$-sparse signals,
$K$-sparse $\alpha$-strongly-decaying $K$-sparse signals,
and $K$-sparse signals with i.i.d. $\mathcal{N}(0, \sigma^2)$ entries for any $\sigma$,
leading to lower bounds for these three classes of sparse signals.

This paper investigates the exact recovery of $K$-sparse signals $\x$ with OMP
in the noise-free case. In practical applications, the observation vector $\y$
in \eqref{e:model} is frequently corrupted with a noise $\v$.
Hence, it is interesting to study the exact support recovery of  $K$-sparse
signals $\x$ with OMP in the noisy case.
Although we cannot directly use the techniques developed in this paper to
characterise the probability of exact support recovery with OMP in the noisy case
since characterizing the probability that $\min_{i\in \Omega}|x_i|$
is not smaller than a term involving the noise vector $\v$ is needed,
we believe that our techniques are useful and can be modified to
lower bound this probability with some new techniques.

In addition to the OMP algorithm, there are many other sparse recovery algorithms,
such as the generalized OMP \cite{WanS12}, basis pursuit \cite{CanT05},
subspace pursuit \cite{DaiM09}, iterative hard thresholding \cite{BluD09} and
Compressive Sampling Matching Pursuit (CoSaMP) \cite{NeeT09},
whether the techniques developed in this paper can be employed to improve
the performance of these algorithms is left as future work.

\section*{ACKNOWLEDGMENT}
We are grateful to the Associate Editor Prof. Michael Wakin for his valuable and thoughtful suggestions.

\appendices
\section{Proof of Lemma~\ref{l:decaying}}
\label{ss:pfldecay}
\begin{IEEEproof}
If $\x=\0$, then it is easy to see that the lemma holds.
In the following, we assume $\x\neq\0$.

Since $\x$ is a $K$-sparse $\alpha$-strongly-decaying signal, by Definition \ref{d:alphastr},
one can see that $\x_{\Omega\setminus \mathcal{S}}$ is also a $|\Omega\setminus \mathcal{S}|$-sparse
$\alpha$-strongly-decaying signal for any $\mathcal{S}\subseteq \Omega$.
Thus, to show the lemma, it suffices to show
\begin{align}
\label{e:x12phi}
\frac{\|\x\|_{1}^2}{\|\x\|_{2}^2}\leq \phi(\ell),
\end{align}
where $\ell$ is the number of nonzero entries of $\x$.

Recall that we assume $K$-sparse $\x$ satisfies \eqref{e:order} throughout this paper,
hence $x_{\ell}\neq0$ and $x_j=0$ for $\ell+1\leq j\leq n$. Furthermore,
by Definition \ref{d:alphastr}, $|x_i|\geq \alpha^{\ell-i}|x_\ell|$ for $1\leq i\leq \ell$.
If $|x_i|=\alpha^{\ell-i}|x_\ell|$ for $1\leq i\leq \ell$, then
\[
\frac{\|\x\|_{1}^2}{\|\x\|_{2}^2}
=\frac{(\sum_{i=1}^{\ell}|x_i|)^2}{\sum_{i=1}^{\ell}|x_i|^2}
=\frac{(\sum_{i=1}^{\ell}\alpha^{\ell-i})^2|x_\ell|^2}
{(\sum_{i=1}^{\ell}\alpha^{2\ell-2i})|x_\ell|^2}=\phi(\ell).
\]
Therefore, \eqref{e:x12phi} holds in this case.
If there exists some $1\leq i\leq \ell-1$ such that $|x_i|>\alpha^{\ell-i}|x_\ell|$,
then by \cite[Lemma 7]{WenZLLT19}, one can see that
$\frac{\|\x\|_{1}^2}{\|\x\|_{2}^2}\leq \phi(\ell)$. Therefore, Lemma \ref{l:decaying} holds.
\end{IEEEproof}

\section{Proof of Lemma~\ref{l:ratio}}
\label{ss:pflratio}
\begin{IEEEproof}
We first show \eqref{e:ratio} holds. Let
\beq
\label{e:w}
w=\frac{\|\u\|_1^2}{\|\u\|_2^2}, \,\;\bar{\u}=\frac{\u}{\|\u\|_2},
\eeq
then $w=\|\bar{\u}\|_1^2$ and $\bar{\u}$ is uniformly distributed
on the surface of $p$-dimensional unit sphere $\mathbb{S}^{p}$.

By \cite[Theorem 2.49]{Fol84}, we have the following equality for the invariant measure on $\mathbb{S}^{p-1}$
\begin{align}
\label{eq:key equality}
\frac{\int_{\mathbb{R}^{p}}f(\bxi)\mathrm{d}\bxi}{\int_{\mathbb{R}^{p}}e^{-\frac{\|\bxi\|_{2}^{2}}{2}}\mathrm{d}\bxi}
=\frac{\int_{\mathbb{S}^{p-1}}\int_{0}^{\infty}f(t\bxi^{\prime})t^{p-1}\mathrm{d}t\mathrm{d}\sigma(\bxi^{\prime})}
{\int_{0}^{\infty}e^{-\frac{t^{2}}{2}}t^{p-1}\mathrm{d}t},
\end{align}
where $\sigma$ is the normalized rotational invariant measure on $\mathbb{S}^{p-1}$,
$\bxi^{\prime}=\frac{\bxi}{\|\bxi\|_2}$ and $t=\|\bxi\|_2$.
Taking
\beq
\label{e:f}
f(\bxi)=\|\bxi\|_{1}^{m}e^{-\frac{\|\bxi\|_{2}^{2}}{2}}
\eeq
for certain positive number $m$,  we have
\begin{align}
\label{e:rhs}
&\,\frac{\int_{\mathbb{S}^{p-1}}\int_{0}^{\infty}f(t\bxi^{\prime})t^{p-1}\mathrm{d}t\mathrm{d}\sigma(\bxi^{\prime})}
{\int_{0}^{\infty}e^{-\frac{t^{2}}{2}}t^{p-1}\mathrm{d}t}\nonumber\\
=&\frac{\int_{\mathbb{S}^{p-1}}\int_{0}^{\infty}\|\bxi^{\prime}\|_{1}^{m}e^{-\frac{t^{2}}{2}}t^{m+p-1}\mathrm{d}t\mathrm{d}\sigma(\bxi^{\prime})}
{\int_{0}^{\infty}e^{-\frac{t^{2}}{2}}t^{p-1}\mathrm{d}t}\nonumber\\
=&\frac{\int_{0}^{\infty}e^{-\frac{t^{2}}{2}}t^{m+p-1}\mathrm{d}t}{\int_{0}^{\infty}e^{-\frac{t^{2}}{2}}t^{p-1}\mathrm{d}t}
\int_{\mathbb{S}^{p-1}}\|\bxi^{\prime}\|_{1}^{m}\mathrm{d}\sigma(\bxi^{\prime})\nonumber\\
=&2^{m/2}\frac{\Gamma((m+p)/2)}{\Gamma(p/2)}\cdot\mathbb{E}[w^{m/2}],
\end{align}
where the last equality follows from the definition of Gamma function,
\eqref{e:w} and $\bxi^{\prime}=\frac{\bxi}{\|\bxi\|_2}$.

By some simple calculations, one can check that, for any given $m>0$,
function $g(\eta)=\eta^m e^{-\eta}, \eta>0$ achieves the maximal value at $\eta=m$.
Hence, $\eta^m\leq m^{m}e^{-m}e^{\eta}$ for any $\eta>0$.
Let $\eta=\|\bxi\|_{1}$, then we have
\[
\|\bxi\|_{1}^{m}\leq m^{m}e^{-m}e^{\|\bxi\|_{1}},
\]
thus by \eqref{eq:key equality} and \eqref{e:f}, we have
\begin{align}
\label{e:lhs}
\frac{\int_{\mathbb{R}^{p}}f(\bxi)\mathrm{d}\bxi}{\int_{\mathbb{R}^{p}}e^{-\frac{\|\bxi\|_{2}^{2}}{2}}\mathrm{d}\bxi}
\leq& m^{m}e^{-m}\frac{\int_{\mathbb{R}^{p}}e^{(\|\bxi\|_{1}-\frac{\|\bxi\|_{2}^{2}}{2})}\mathrm{d}\bxi}
{\int_{\mathbb{R}^{p}}e^{-\frac{\|\bxi\|_{2}^{2}}{2}}\mathrm{d}\bxi}\nonumber\\
=&m^{m}e^{-m}\left(\frac{\int_{-\infty}^{\infty}e^{(|s|-\frac{s^{2}}{2})}\mathrm{d}s}{\int_{-\infty}^{\infty}e^{-\frac{s^{2}}{2}}\mathrm{d}s}\right)^{p}\nonumber\\
\leq &2.775^{p}m^{m}e^{-m}.
\end{align}

By \eqref{eq:key equality}, \eqref{e:rhs} and \eqref{e:lhs}, we have
\begin{align*}
\mathbb{E}[w^{m/2}]& \leq m^{m}e^{-m}\frac{2.775^{p}}{2^{m/2}}\frac{\Gamma(p/2)}{\Gamma((m+p)/2)}.
\end{align*}
By \cite[Theorem 1]{Jam15}, for $\forall s>0$, it holds that
\[
\sqrt{2\pi}s^{s-1/2}e^{-s}\leq \Gamma(s)\leq \sqrt{2\pi}s^{s-1/2}e^{-s}e^{1/(12s)}.
\]
Hence,
\begin{align*}
\mathbb{E}[w^{m/2}]& \leq m^{m}e^{-m}\frac{2.775^{p}}{2^{m/2}}
\frac{(\frac{p}{2})^{\frac{p-1}{2}}e^{-\frac{p}{2}}e^{\frac{1}{6p}}}{(\frac{m+p}{2})^{\frac{m+p-1}{2}}e^{-\frac{m+p}{2}}}\\
& =2.775^{p} m^{m}e^{-\frac{m}{2}}\frac{p^{\frac{p}{2}}}{(m+p)^{\frac{m+p}{2}}}\sqrt{1+\frac{m}{p}}e^{\frac{1}{6p}}.
\end{align*}
By using the Markov inequality, the fact that $p\geq 1$ and let $\gamma=m/p$, we have
\begin{align}
\label{e:lhslbd}
\mathbb{P}(w\geq\mu p)= & \mathbb{P}(w^{m/2}\geq(\mu p)^{m/2})\leq\frac{\mathbb{E}[w^{m/2}]}{(\mu p)^{m/2}} \nonumber\\
\leq & \left(\frac{2.775\gamma^{\gamma}e^{-\frac{\gamma}{2}}(1+\gamma)^{-\frac{1+\gamma}{2}}}{\mu^{\gamma/2}}\right)^{p}\sqrt{1+\gamma}e^{\frac{1}{6}}.
\end{align}
By  \eqref{e:w} and \eqref{e:lhslbd}, one can easily see that \eqref{e:ratio} holds.

We then show \eqref{e:ratio1} holds. Let $\mu=0.95$ and $\gamma=1.505$, then by \eqref{e:lhslbd}, we have
\[
\mathbb{P}(w\geq0.95p)\leq1.87\times0.796^{p}.
\]
Then \eqref{e:ratio1} follows from \eqref{e:w}.
\end{IEEEproof}

\section{Proof of Lemma~\ref{l:Gaussian}}
\label{ss:Gaussian}

Before proving Lemma \ref{l:Gaussian}, we need to introduce the following lemma:

\begin{lemma}
\label{l:bdonC}
Let integers $n$ and $p$ satisfy $n/2< p< n$, then
\begin{align*}
\sum_{i=p}^{n} {n \choose i} < \frac{p}{2p-n}{n \choose p}.
\end{align*}
\end{lemma}

\begin{IEEEproof}
It is not difficult to check that
\begin{align*}
\sum_{i=p}^{n} {n \choose i} &= {n \choose p} \sum_{i=p}^{n}
\left(\frac{p!(n-p)!}{i!(n-i)!}\right)\\
&= {n \choose p}
\left(1+\sum_{i=p+1}^{n}\frac{(n-i+1)\cdots (n-p)}{(p+1)\cdots i}\right)\\
&\leq {n \choose p}
\left[1+\sum_{i=p+1}^{n}\left(\frac{n-p}{p+1}\right)^{i-p}\right]\\
&< {n \choose p}
\left[\sum_{j=0}^{\infty}\left(\frac{n-p}{p+1}\right)^{j}\right]\\
&= {n \choose p}\frac{p+1}{2p+1-n}\\
&< {n \choose p}\frac{p}{2p-n},
\end{align*}
where the last equality follows from the assumption that $n/2< p< n$.
\end{IEEEproof}

In the following, we prove Lemma~\ref{l:Gaussian}.

\begin{IEEEproof}
By \eqref{e:phit2} and the Cauchy-Schwarz inequality, one can see that, for any
$\mathcal{S}\subseteq \Omega$ with $|\mathcal{S}|\leq \lceil0.95K\rceil$, it holds that
\[
\|\x_{\mathcal{S}}\|_{1}^2\leq \phi(|\mathcal{S}|)
\|\x_{\mathcal{S}}\|_{2}^2.
\]
Since $\x$ is $K$-sparse, $|\Omega|\leq K$.
If $|\Omega|\leq\lceil0.95K\rceil$, then by the above inequality, \eqref{e:pGaussian} holds.
In the following, we assume $|\Omega|> \lceil0.95K\rceil$.
Then to prove \eqref{e:pGaussian}, it suffices to show that
\beq
\label{e:Gauss1}
\mathbb{P}(\bigcup_{|\mathcal{S}|=\lceil0.95K\rceil}^{|\Omega|}
(\frac{\|\x_{\mathcal{S}}\|_{1}^2}{\|\x_{\mathcal{S}}\|_{2}^2}> \phi(|\mathcal{S}|)))
\leq \frac{3.614}{\sqrt{K}}\times0.981^{\lceil0.95K\rceil}.
\eeq

By some fundamental calculations, we have
\begin{align}
\mathbb{P}&
(\bigcup_{|\mathcal{S}|=\lceil0.95K\rceil}^{|\Omega|}
(\|\x_{\mathcal{S}}\|_{1}^2> \phi(|\mathcal{S}|)
\|\x_{\mathcal{S}}\|_{2}^2))\nonumber\\
&\leq
\sum_{|\mathcal{S}|=\lceil0.95K\rceil}^{|\Omega|}
\mathbb{P}(\|\x_{\mathcal{S}}\|_{1}^2> \phi(|\mathcal{S}|)\|\x_{\mathcal{S}}\|_{2}^2)\nonumber\\
&\overset{(a)}{\leq}
\sum_{|\mathcal{S}|=\lceil0.95K\rceil}^{|\Omega|}
\left[{|\Omega| \choose |\mathcal{S}|}\times1.87\times0.796^{|\mathcal{S}|}\right]\nonumber\\
&\leq
1.87\times 0.796^{\lceil0.95K\rceil}\times
\sum_{|\mathcal{S}|=\lceil0.95K\rceil}^{|\Omega|}{|\Omega| \choose |\mathcal{S}|}\nonumber\\
&\overset{(b)}{\leq}
1.87\times 0.796^{\lceil0.95K\rceil}\times
\frac{\lceil0.95K\rceil}{2\lceil0.95K\rceil-|\Omega|}\times{|\Omega| \choose \lceil0.95K\rceil}\nonumber\\
&\leq
1.87\times 0.796^{\lceil0.95K\rceil}\times\frac{0.95K}{1.9K-K}\times{|\Omega| \choose \lceil0.95K\rceil}\nonumber\\
&<
1.974\times 0.796^{\lceil0.95K\rceil}\times{K \choose \lceil0.95K\rceil},
\label{e:Gauss11}
\end{align}
where (a) and (b) respectively follow from \eqref{e:ratio1} and Lemma \ref{l:bdonC}.

In the following, we upper bound ${K \choose \lceil0.95K\rceil}$.
By \cite{Rob95}, for any integer $n$, we have
\[
\sqrt{2\pi}n^{n+1/2}e^{-n}e^{\frac{1}{12n+1}}<n!<\sqrt{2\pi}n^{n+1/2}e^{-n}e^{\frac{1}{12n}}.
\]
Since $\lceil0.95K\rceil< K$, we have
\begin{align}
&{K\choose \lceil0.95K\rceil} \nonumber\\
&\hspace{2mm} = \frac{K!}{\lceil0.95K\rceil!(K-\lceil0.95K\rceil)!}
\nonumber\\
&\hspace{2mm} <\frac{\sqrt{2\pi}e^{-K}e^{\frac{1}{12K}}}
{2\pi e^{-K}e^{\frac{1}{12\lceil0.95K\rceil+1}
+\frac{1}{12(K-\lceil0.95K\rceil)+1}}}\nonumber\\
&\hspace{2mm} \times\frac{K^{K+1/2}}
{\lceil0.95K\rceil^{\lceil0.95K\rceil+1/2}
(K-\lceil0.95K\rceil)^{K-\lceil0.95K\rceil+1/2}}\nonumber\\
&\hspace{2mm} <\frac{K^{K+1/2}}{\sqrt{2\pi}
\lceil0.95K\rceil^{\lceil0.95K\rceil+\frac{1}{2}}
(K-\lceil0.95K\rceil)^{K-\lceil0.95K\rceil+\frac{1}{2}}}\nonumber\\
&\hspace{2mm} \overset{(a)}{\leq}\frac{K^{K+1/2}}{\sqrt{2\pi}
(0.95K)^{0.95K+1/2}(K-0.95K)^{K-0.95K+1/2}}
\nonumber\\
&\hspace{2mm} =\frac{K^{K+1/2}}{\sqrt{2\pi}
(0.95K)^{0.95K+1/2}(0.05K)^{0.05K+1/2}}\nonumber\\
&\hspace{2mm} =\frac{1}{\sqrt{2\pi K}}
\frac{1}{(0.95)^{0.95K+1/2}(0.05)^{0.05K+1/2}}\nonumber\\
&\hspace{2mm} =\frac{1}{\sqrt{2\pi K}}(0.95\times 0.05)^{-1/2}
\left(0.95^{-1}0.05^{-\frac{0.05}{0.95}}\right)^{0.95K}\nonumber\\
&\hspace{2mm} <\frac{1.8305}{\sqrt{K}}1.2324^{\lceil0.95K\rceil},
\label{e:Gauss12}
\end{align}
where (a) is because $s^{s+1/2}(K-s)^{K-s+1/2}$
is an increase function of $s$ for $K/2\leq s<K$.
Hence, \eqref{e:Gauss1} follows from \eqref{e:Gauss11} and \eqref{e:Gauss12}.
\end{IEEEproof}

\section{Comparison of the lower bounds on $m$ given by  Corollary \ref{c:nbofob} and
 \cite[Corollary 7]{TroG07b} }
\label{ss:comp}

The proof of \cite[Corollary 7]{TroG07b}  shows that for any fixed $K, n$ and $\delta\in (0,0.36)$,
if $m$ satisfy
\beq
\label{e:mtrop}
m\geq\left(2+\sqrt{\frac{1}{\ln(n/\delta)}}+\sqrt{\frac{2}{K}}\right)^2K\ln(n/\delta),
\eeq
$\A$ is being defined in Theorem \ref{t:nbofob} and $\x$ is a $K$-sparse signal. Then
\beq
\label{e:pblbtrop}
\mathbb{P}(\mathbb{S})\geq1-\left(\frac{\delta}{n}+\frac{\delta^2}{4}\right),
\eeq
where event $\mathbb{S}$ is defined in \eqref{e:E}.

In the following, we show that the requirement on $m$ given by Corollary \ref{c:nbofob} is weaker
than that given by \cite[Corollary 7]{TroG07b}. To this end, we assume
$
\zeta=\frac{\delta}{n}+\frac{\delta^2}{4},
$
then
\beq
\label{e:deltatrop}
\delta=\frac{2(\sqrt{1+n^2\zeta}-1)}{n}.
\eeq
Hence, \eqref{e:mtrop} is equivalent to
\begin{align}
\label{e:mtrop2}
m&\geq\left(2\sqrt{K\ln(n/\delta)}+\sqrt{K}+\sqrt{2\ln(n/\delta)}\right)^2\nonumber\\
&=\left((2\sqrt{K}+\sqrt{2})\sqrt{\ln\frac{n^2}{2(\sqrt{1+n^2\zeta}-1)}}+\sqrt{K}\right)^2\nonumber\\
&=\left((2\sqrt{K}+\sqrt{2})\sqrt{\ln\frac{\sqrt{1+n^2\zeta}+1}{2\zeta}}+\sqrt{K}\right)^2.
\end{align}
Therefore, \cite[Corollary 7]{TroG07b}  can be equivalently stated as that if $m$ satisfies \eqref{e:mtrop2},
then \eqref{e:prob31} holds

Since $\delta=\frac{n\sqrt{\pi}\zeta}{n+\sqrt{\pi}}$ in Corollary \ref{c:nbofob},
similar to the derivation of \eqref{e:mtrop2}, \eqref{e:m2} can be  rewritten as
\beq
\label{e:mequivalent}
m\geq\left((\sqrt{2\beta K}+\sqrt{2})\sqrt{\ln\frac{n+\sqrt{\pi}}{\sqrt{\pi}\zeta}}+\sqrt{K}\right)^2.
\eeq

By the above analysis, to show that the requirement on $m$ given by Corollary \ref{c:nbofob} is weaker
than that given by \cite[Corollary 7]{TroG07b}, it is equivalent to show that
\begin{multline}
\label{e:compm}
(2\sqrt{K}+\sqrt{2})\sqrt{\ln\frac{\sqrt{1+n^2\zeta}+1}{2\zeta}}\\
>(\sqrt{2\beta K}+\sqrt{2})\sqrt{\ln\frac{n+\sqrt{\pi}}{\sqrt{\pi}\zeta}}
\end{multline}
Since $\sqrt{\ln t}$ changes very slowly as $t$ changes when $t$ is large (say larger than 10),
the left-hand side and right-hand side of \eqref{e:compm} are respectively dominated by
$2\sqrt{K}$ and $\sqrt{2\beta K}$. Thus to show \eqref{e:compm}, we only show $\beta<2$.
Since in Corollary \ref{c:nbofob} $\delta=\frac{n\sqrt{\pi}\zeta}{n+\sqrt{\pi}}$ and
$\zeta\leq \frac{1}{\sqrt{\pi}}$, $\ln(n/\delta)>1$. Therefore, by \eqref{e:alpha2}, to show $\beta<2$,
it suffices to show that
\[
2n^2\geq (n-K)(K+1).
\]
Since $1\leq K\leq n$, the above inequality holds.

\section{Proof of Lemma \ref{l:main} }
\label{ss:pflmain}

\begin{IEEEproof}
Since $\mathcal{S}_k\subseteq \Omega$ and $|\mathcal{S}_k|=k$,
$\mathcal{S}_{k+1}\subseteq \Omega$ and $|\mathcal{S}_{k+1}|=k+1$ if and only if
$s^{k+1}\in \Omega\setminus \mathcal{S}_k$.
Thus, by line 2 of Algorithm \ref{a:OMP}, to show the lemma, it suffices to show that
\beq
\label{e:l2condoriginal}
\max_{i\in \Omega}|\langle \rr^{k},\A_i\rangle|
> \max_{j\in \Omega^c}|\langle \rr^{k},\A_j\rangle|.
\eeq
Indeed, if \eqref{e:l2condoriginal} holds, then $s^{k+1}\in \Omega$.
Moreover, by lines 4 and 5 of Algorithm \ref{a:OMP}, we have $|\langle \rr^{k},\A_i\rangle|=0$
for $i\in \mathcal{S}_k$.
Hence, by \eqref{e:l2condoriginal}, $s^{k+1}\in \Omega\setminus \mathcal{S}_k$.

By  \cite[(31-38)]{WenZWM17} with the noise vector $\v=\0$,
one can see that \eqref{e:l2condoriginal} holds if the following inequality holds:
\begin{align}
\|\A^\top_{\Omega\setminus \mathcal{S}_k}\P^{\perp}_{\mathcal{S}_k}\A_{\Omega\setminus \mathcal{S}_k}\x_{\Omega\setminus \mathcal{S}_k}\|_{\infty}
>\|\A^\top_{\Omega^c}\P^{\perp}_{\mathcal{S}_k}\A_{\Omega\setminus \mathcal{S}_k}\x_{\Omega\setminus \mathcal{S}_k}\|_{\infty}
\label{e:l2cond1}.
\end{align}

In the following, we give a lower bound on the left-hand side of \eqref{e:l2cond1}.
Since the inequality in \eqref{e:csk} holds with $\mathcal{S}=\Omega\setminus\mathcal{S}_k$,
by \cite[(3.1), (3.3) and (4.1)]{WenZLLT19},
we have
\[
\|\A^\top_{\Omega\setminus \mathcal{S}_k}\P^{\perp}_{\mathcal{S}_k}\A_{\Omega\setminus \mathcal{S}_k}\x_{\Omega\setminus \mathcal{S}_k}\|_{\infty}
\geq\frac{\|\P^{\perp}_{\mathcal{S}_k}\A_{\Omega\setminus \mathcal{S}_k}\x_{\Omega\setminus \mathcal{S}_k}\|_{2}^2}{\sqrt{\phi(|\Omega\setminus \mathcal{S}_k|)}
\|\x_{\Omega\setminus \mathcal{S}_k}\|_{2}}.
\]
Since $\x$ is $K$-sparse, $\mathcal{S}_k\subseteq \Omega$ and $|\mathcal{S}_k|=k$,
$|\Omega\setminus \mathcal{S}_k|\leq K-k$. Since $\phi(t)$ is a nondecreasing function,
by the above inequality, we have
\begin{align}
\label{e:linflbd}
\|\A^\top_{\Omega\setminus \mathcal{S}_k}\P^{\perp}_{\mathcal{S}_k}\A_{\Omega\setminus \mathcal{S}_k}\x_{\Omega\setminus \mathcal{S}_k}\|_{\infty}
\geq\frac{\|\P^{\perp}_{\mathcal{S}_k}\A_{\Omega\setminus \mathcal{S}_k}\x_{\Omega\setminus \mathcal{S}_k}\|_{2}^2}{\sqrt{\phi(K-k)}\|\x_{\Omega\setminus \mathcal{S}_k}\|_{2}}.
\end{align}
Let $\bar{\sigma}$ denote the smallest singular value of
$\P^{\perp}_{\mathcal{S}_k}\A_{\Omega\setminus \mathcal{S}_k}$,
then by \cite[Lemma 5]{CaiW11}, we have $\bar{\sigma}\geq \sigma_{\min}$, hence
\[
\|\P^{\perp}_{\mathcal{S}_k}\A_{\Omega\setminus \mathcal{S}_k}\x_{\Omega\setminus \mathcal{S}_k}\|_{2}
\geq \bar{\sigma}\|\x_{\Omega\setminus \mathcal{S}_k}\|_{2}
\geq \sigma_{\min}\|\x_{\Omega\setminus \mathcal{S}_k}\|_{2}.
\]
Then, by \eqref{e:uk} and \eqref{e:linflbd}, we have
\begin{align*}
&\|\A^\top_{\Omega\setminus \mathcal{S}_k}\P^{\perp}_{\mathcal{S}_k}\A_{\Omega\setminus \mathcal{S}_k}\x_{\Omega\setminus \mathcal{S}_k}\|_{\infty}
-\|\A^\top_{\Omega^c}\P^{\perp}_{\mathcal{S}_k}\A_{\Omega\setminus \mathcal{S}_k}\x_{\Omega\setminus \mathcal{S}_k}\|_{\infty} \nonumber\\
&\hspace{2mm} \geq\|\P^{\perp}_{\mathcal{S}_k}\A_{\Omega\setminus \mathcal{S}_k}\x_{\Omega\setminus \mathcal{S}_k}\|_{2}\nonumber\\
&\hspace{2mm} \times\left(\frac{\|\P^{\perp}_{\mathcal{S}_k}\A_{\Omega\setminus \mathcal{S}_k}\x_{\Omega\setminus \mathcal{S}_k}\|_{2}}{\sqrt{\phi(K-k)}\|\x_{\Omega\setminus \mathcal{S}_k}\|_{2}}
-\|\A^\top_{\Omega^c}\u_k\|_{\infty}\right)\nonumber\\
&\hspace{2mm} \geq\sigma_{\min}\|\x_{\Omega\setminus \mathcal{S}_k}\|_{2}
\left(\frac{\sigma_{\min}}{\sqrt{\phi(K-k)}}
-\|\A^\top_{\Omega^c}\u_k\|_{\infty}\right).
\end{align*}
Then, one can easily see that \eqref{e:l2cond1} holds if \eqref{e:main} holds,
and hence the lemma holds.
\end{IEEEproof}

\section{Proof of Lemma \ref{l:Gaussian1} }
\label{ss:pflGauss}

\begin{IEEEproof}
One can check that
\begin{align*}
\mathbb{P}&(\bigcap_{i=1}^\ell(\|\B^{\top}\u_i\|_{\infty}\leq\epsilon_i))\\
&=\,\mathbb{P}(\|\B^{\top}\u_1\|_{\infty}\leq\epsilon_1,\ldots, \|\B^{\top}\u_\ell\|_{\infty}\leq\epsilon_\ell)\\
&=\,\mathbb{P}(|\B_1^{\top}\u_1|\leq\epsilon_1,\ldots,
|\B_p^{\top}\u_1|\leq\epsilon_1,\ldots, \\
&\;\;\;\quad|\B_1^{\top}\u_\ell|\leq\epsilon_\ell,\ldots,
|\B_p^{\top}\u_\ell|\leq\epsilon_\ell)\\
&\geq\,\mathbb{P}(|\B_1^{\top}\u_1|\leq\epsilon_1)\times\cdots\times
\mathbb{P}(|\B_p^{\top}\u_1|\leq\epsilon_1)\times\cdots\times \\
&\quad\,\;\mathbb{P}(|\B_1^{\top}\u_\ell|\leq\epsilon_\ell)\times\cdots\times
\mathbb{P}(|\B_p^{\top}\u_\ell|\leq\epsilon_\ell)\\
&=\,\prod_{i=1}^{\ell}\prod_{j=1}^p\mathbb{P}(|\B_j^{\top}\u_i|\leq\epsilon_i),
\end{align*}
where the inequality follows from \cite[Theorem 1]{Sidak67}.

Since the columns of $\B$ are independent and $\B$ is independent with $\u_i$ for $1\leq i\leq \ell$,
to show \eqref{e:pbd}, we only need to prove that for any $1\leq i\leq \ell, 1\leq j\leq p$,
it holds that
\beq
\label{e:err}
\mathbb{P}(|\B_j^{\top}\u_i|\leq\epsilon_i)\geq
1-\frac{e^{-\epsilon_i^2m/2}}{\sqrt{\pi m/2}\epsilon_i}.
\eeq

We first prove that \eqref{e:err} holds for $\|\u_i\|_2= 1$.
Since the entries of $\B$ independently and identically follow
$\mathcal{N}(0, 1/m)$ distribution and $\|\u_i\|_2= 1$,
$\B_j^{\top}\u_i\sim\mathcal{N}(0, 1/m)$. Thus
\begin{align*}
\mathbb{P}(|\B_j^{\top}\u_i|\leq\epsilon_i)=&
\frac{1}{\sqrt{2\pi/m}}\int_{-\epsilon_i}^{\epsilon_i}e^{-m\eta^2/2}d\eta \nonumber\\
=&1-\frac{2}{\sqrt{2\pi/m}}\int_{\epsilon_i}^{\infty}e^{-m\eta^2/2}d\eta
\nonumber\\
\overset{(a)}{=}&1-\frac{2}{\sqrt{\pi}}\int_{\sqrt{m/2}\epsilon_i}^{\infty}
e^{-\xi^2}d\xi\nonumber\\
\overset{(b)}{\geq}&1-\frac{e^{-\epsilon_i^2m/2}}{\sqrt{\pi m/2}\epsilon_i},
\end{align*}
where (a) follows from the integral transformation and (b) is from \cite[(4)]{KarL07}.
Thus, \eqref{e:err} holds for $\|\u_i\|_2= 1$.

We then prove \eqref{e:err} holds for $\|\u_i\|_2<1$.
Since $\|\u_i\|_2<1$, we have
\begin{align*}
\mathbb{P}(|\B_j^{\top}\u_i|\leq\epsilon_i)=&
\mathbb{P}(|\B_j^{\top}(\u_i/\|\u_i\|_2)|\leq\epsilon_i/\|\u_i\|_2)\\
\geq &\mathbb{P}(|\B_j^{\top}(\u_i/\|\u_i\|_2)|\leq\epsilon_i)\\
\geq &1-\frac{e^{-\epsilon_i^2m/2}}{\sqrt{\pi m/2}\epsilon_i}.
\end{align*}
Thus, \eqref{e:err} holds for $\|\u_i\|_2<1$.
\end{IEEEproof}

\section{Proof of Lemma \ref{l:varphi} }
\label{ss:varphi}

\begin{IEEEproof}
Let
\[
\bar{\varphi}(t)=\frac{1}{\sqrt{t}e^{1/t}},\,\; 0<t<2,
\]
then by some direct calculations, we have
\[
\bar{\varphi}'(t)=\frac{2-t}{2t^{5/2}e^{1/t}}>0.
\]
Since $0<t_1\leq t_2\leq \ldots\leq t_p<2$, we have
\begin{align*}
\sum_{i=1}^p\varphi(t_i)=\sum_{i=1}^p(t_i\bar{\varphi}(t_i))\leq\sum_{i=1}^p(t_i\bar{\varphi}(t_p))
= \frac{\sum_{i=1}^pt_i}{\sqrt{t_p}e^{1/t_p}}.
\end{align*}
Hence \eqref{e:varphi2} holds.
\end{IEEEproof}

\bibliographystyle{IEEEtran}
\bibliography{ref}

\end{document}

%% file: micros.tex
\usepackage{amsmath}
\usepackage{amsfonts}
\usepackage{amssymb}
\usepackage{algorithm}
\usepackage{algorithmic}
\usepackage{graphicx}
\usepackage{verbatim}

\newtheorem{theorem}{Theorem}
\newtheorem{lemma}{Lemma}
\newtheorem{remark}{Remark}

\newtheorem{corollary}{Corollary}
\newtheorem{example}{Example}
\newtheorem{definition}{Definition}

\newcommand{\beq}{\begin{equation}}
\newcommand{\eeq}{\end{equation}}
\newcommand{\beqnn}{\begin{equation*}}
\newcommand{\eeqnn}{\end{equation*}}
\newcommand{\beqy}{\begin{eqnarray}}
\newcommand{\eeqy}{\end{eqnarray}}
\newcommand{\beqynn}{\begin{eqnarray*}}
\newcommand{\eeqynn}{\end{eqnarray*}}
\newcommand{\bit}{\begin{itemize}}
\newcommand{\eit}{\end{itemize}}
\newcommand{\ben}{\begin{enumerate}}
\newcommand{\een}{\end{enumerate}}
\newcommand{\bex}{\begin{example}}
\newcommand{\eex}{\end{example}}

\newcommand{\balg}[1]{\begin{algorithm} \caption{#1}}
\newcommand{\ealg}{\end{algorithm}}

\newcommand{\balgc}{\begin{algorithmic}[1]}
\newcommand{\ealgc}{\end{algorithmic}}

\newcommand{\bary}{\begin{array}}
\newcommand{\eary}{\end{array}}
\newcommand{\bmx}{\begin{bmatrix}}
\newcommand{\emx}{\end{bmatrix}}
\newcommand{\bsmx}{\left[\begin{smallmatrix}}
\newcommand{\esmx}{\end{smallmatrix}\right]}
\newcommand{\bmxc}[1]{\left[\begin{array}{@{}#1@{}}}
\newcommand{\emxc}{\end{array}\right]}
\newcommand{\bcn}{\begin{center}}
\newcommand{\ecn}{\end{center}}

\newcommand{\A}{\boldsymbol{A}}
\newcommand{\B}{\boldsymbol{B}}

\newcommand{\I}{\boldsymbol{I}}

\renewcommand{\P}{\boldsymbol{P}}

\renewcommand{\S}{\boldsymbol{S}}

\newcommand{\e}{\boldsymbol{e}}

\newcommand{\rr}{\boldsymbol{r}}

\renewcommand{\u}{\boldsymbol{u}}
\renewcommand{\v}{\boldsymbol{v}}

\newcommand{\x}{{\boldsymbol{x}}}
\newcommand{\y}{{\boldsymbol{y}}}
\newcommand{\z}{\boldsymbol{z}}
\newcommand{\0}{{\boldsymbol{0}}}

\newcommand{\bxi}{\boldsymbol{\xi}}
